Open-Source Framework for the Visualization, Quantification, and Analysis of 4D Flow and High-Dimensional Magnetic Resonance Images

Andras Lasso[*,a], Julia Iacovella[*,b], Csaba Pinter[c], Kevin K. Whitehead[d,e], Sehdev Pressley[b], Danish Vaiyani[d], Matthew Daemer[b], Katherine Hurm[b], Analise M. Sulentic[b], Christian Herz[b], Ryan T. Black[b], Seda Aslan[b], Christopher W. Roy[f], Mark A. Fogel[d,e], Jing Huang[g,h], Matthew A. Jolley[b,d].

*These authors contributed equally to this work.*

**Affiliations**

a Laboratory for Percutaneous Surgery, School of Computing, Queen's University, Kingston, ON, Canada

b Department of Anesthesiology and Critical Care Medicine, Children's Hospital of Philadelphia, 3401 Civic Center Blvd, Philadelphia, PA, USA.

c Ebatinca S.L., Las Palmas de Gran Canaria, Spain. Laboratory for Percutaneous Surgery, School of Computing, Queen's University, Kingston, ON, Canada.

d Division of Cardiology, Children's Hospital of Philadelphia, 3401 Civic Center Blvd, Philadelphia, PA, USA.

e Department of Radiology, Children's Hospital of Philadelphia, Philadelphia, PA, USA.

f Department of Diagnostic and Interventional Radiology, Lausanne University Hospital (CHUV) and University of Lausanne, Lausanne, Switzerland.

g Department of Biostatistics, Epidemiology, and Informatics, Perelman School of Medicine, University of Pennsylvania, Philadelphia, PA, USA

h Department of Pediatrics, Perelman School of Medicine, The University of Pennsylvania PolicyLab, Children's Hospital of Philadelphia, Philadelphia, PA, USA

*Co-First: These authors contributed equally

Word Count: 6,190

Manuscript Type: Original Research

Funding: This work was supported by NIH 1R01HL153166, A Children's Hospital of Philadelphia Post-Frontier Award, the Cora Topolewski Fund at CHOP Pediatric Valve Center, Additional

Ventures CTIP Device Innovation Award, the Topolewsky Endowed Chain in Pediatric Cardiology, and The Mussallem Congenital Heart Disease Alliance

Data Sharing Statement: All code and data generated or analyzed during the study are available in anonymized format on GitHub.

**Corresponding Author:**

Matthew Jolley

Children’s Hospital of Philadelphia

University of Pennsylvania School of Medicine

3401 Civic Center Blvd

Philadelphia, PA 19104

Phone 215-280-5643

Email: JOLLEYM@chop.edu, Matthew.Jolley@pennmedicine.upenn.edu

Open-Source Framework for the Visualization, Quantification, and Analysis of 4D Flow and High-Dimensional Magnetic Resonance Images

Article Type: Original Research

**Abbreviations:**

2D = two-dimensional

3D = three-dimensional

4D = four-dimensional

CFD = computational fluid dynamics

CHD = congenital heart disease

CMR = cardiovascular magnetic resonance

HDim = high-dimensional

ICC = intraclass correlation coefficient

ITK = Insight Toolkit

NRRD = Nearly Raw Raster Data

## Abstract

**Background:** Four-dimensional (4D) flow cardiovascular magnetic resonance (CMR) captures three-dimensional (3D) anatomy and blood velocity across the cardiac cycle, but current post-processing software is closed, varies across platforms, and sits outside open-source image-computing frameworks. We developed an open-source framework for the import, visualization, and quantification of 4D flow and other high-dimensional (HDim) CMR within SlicerHeart, an extension for 3D Slicer.

**Methods:** The framework comprises two interactive 3D Slicer modules: Import 4D Flow reads vendor Digital Imaging and Communications in Medicine (DICOM), applies preprocessing, and stores velocities as a time-resolved vector field; Analyze 4D Flow provides measurement planes, streamline and Doppler-like visualization, segmentation propagation, and hemodynamic quantification. Underlying both, we extended the Insight Toolkit (ITK) to store five- and higher-dimensional arrays. We compared flow quantification with a commercial package (cvi42) and two-dimensional (2D) phase contrast in 13 patients with congenital heart disease (CHD) across 108 paired regions of interest, using Bland-Altman analysis adjusted for multiple regions per patient and intraclass correlation coefficients (ICCs). We also implemented initial consensus-recommended processing methods as a standalone open-source library with an automated test suite of analytic and computational fluid dynamics (CFD) phantoms and clinical CMR acquisitions.

**Results:** The modules integrate import, preprocessing, visualization, and quantification with 3D Slicer's segmentation, registration, and rendering tools, and export velocity fields and geometry for computational modeling. Agreement with cvi42 was close, with small mean differences: forward volume bias −0.1 mL (95% limits of agreement 2.2 to −2.3; ICC 0.997) and effective volume per minute bias −0.015 L/min (0.21 to −0.24; ICC 0.994). Against 2D phase contrast, forward volume bias (−2.2 mL) and mean absolute percentage error (21.8%) were similar to those of cvi42 4D flow (−2.1 mL; 21.3%).

**Conclusions:** This open-source, extensible framework brings 4D flow and other HDim CMR into an established image-computing ecosystem and agrees closely with a widely used commercial reference. Because its infrastructure and methods are open, every step can be inspected, reproduced, and improved, providing a transparent foundation for hemodynamic analysis and method development, from clinically relevant measurements to computational modeling.

## Keywords

## Introduction

Cardiovascular magnetic resonance (CMR) is central to the noninvasive assessment of congenital and structural heart disease, providing radiation-free quantification of ventricular volumes and function, shunts, pulmonary-to-systemic flow ratios (Qp:Qs), and valvular regurgitation, alongside detailed depiction of complex anatomy and flow [1]. Two-dimensional (2D) phase-contrast imaging remains the reference standard for flow, but four-dimensional (4D) flow now encodes velocity in all three orthogonal spatial directions throughout the cardiac cycle, providing volumetric hemodynamic assessment of the heart and great vessels in a single acquisition [2,3]. In parallel, contrast-enhanced and free-running, motion-resolved whole-heart techniques increasingly yield rich 4D (three-dimensional (3D)+time(4D cine)), 4D flow, five-dimensional (3D+time+flow vectors+respiratory motion (5D)), and six-dimensional (3D+time+flow vectors+respiratory motion+R-R or other physiologic parameter (6D)) datasets of both cardiac structure and flow with high spatial and temporal resolution, with particular utility in the small and complex anatomy of congenital heart disease (CHD) [4–6]. Beyond direct measurement, these high-dimensional (HDim) CMR datasets are increasingly used to inform and validate computational fluid dynamics (CFD) and fluid-structure interaction (FSI) simulations, which are emerging as tools to study hemodynamics and to guide surgical and interventional planning in both structural and congenital heart disease [7–11].

Each HDim CMR study carries a wealth of dynamic structural and hemodynamic information across the cardiac cycle. Much of this potential remains latent, however, because the software required to explore, quantify, and build upon these data has not kept pace with the imaging itself. Realizing the full value of HDim CMR depends on an accessible, extensible, and integrated platform, one in which flow, quantification, and the underlying anatomy can be examined together and readily extended to incorporate new capabilities into novel analyses and visualizations. Most available packages fall short of this; they are closed source and operate largely in isolation from tools for image processing and modeling. Further, they can be costly for research use and yield quantified results that vary meaningfully between platforms, limiting comparability and reproducibility [12]. Successive 4D flow CMR consensus statements likewise identify post-processing and validation as unresolved challenges [3,13].

Recent work has begun to reveal this potential, including new capabilities for visualizing myocardium, valves, and blood flow from 4D cine and 4D flow CMR [14]. Here we extend and integrate these nascent capabilities into an open-source, extensible framework for the import, visualization, and quantification of HDim CMR, spanning both cardiac structure and flow, implemented within SlicerHeart, a cardiac-focused extension of the open-source 3D Slicer platform [15,16]. As a core contribution, we extended the Nearly Raw Raster Data (NRRD) reader and writer of the Insight Toolkit (ITK), the open-source library underlying image processing in 3D Slicer [17], to natively support HDim arrays (three spatial axes, a vector-component axis (flow), time, other), allowing time-resolved image, vector-field, and transform sequences to be stored and interchanged within 3D Slicer and by any ITK-based tool without custom code (Appendix A). Because it is built within the established Slicer and SlicerHeart ecosystem, the framework interoperates directly with mature open-source tools, including machine-learning-based segmentation [18,19], deformable registration [20,21], and immersive virtual and augmented reality [22]. Flow can thus be examined in concert with anatomy and carried from interactive review through quantitative measurement within a single environment, offering synergies a standalone application cannot. We present a preliminary validation of the framework's flow measurements against a widely used commercial package and 2D phase contrast, hypothesizing that the framework would agree closely with the commercial package, and demonstrate the framework in a pediatric cohort, including its potential to provide patient-specific inputs and reference data to inform and validate CFD and FSI simulations.

## Methods

### Study design and population

This retrospective, single-center study was conducted under a protocol (IRB-16-012853) determined to be exempt under 45 CFR 46.104(d)(4)(iii) by the Committees for the Protection of Human Subjects (Institutional Review Board) of the Children's Hospital of Philadelphia, with a waiver of consent and of Health Insurance Portability and Accountability Act (HIPAA) authorization. Patients who had undergone clinically indicated CMR including 4D flow were identified from the institutional imaging database, and those with 4D flow datasets analyzed on both platforms were included (n = 13). Patient age, sex, and cardiovascular diagnoses were recorded (Table 1). This cohort was used both to demonstrate the framework across a range of congenital anatomy and for the quantitative comparison against a commercial reference. Additional cases are shown for illustration only.

**Image acquisition**

CMR was performed on a 1.5 Tesla Siemens Avanto Fit scanner (Siemens Healthcare USA, Malvern, PA). 4D flow was obtained after infusion of 4 mg/kg ferumoxytol (AMAG Pharmaceuticals, Waltham, MA) or 20 minutes after injection of 0.2 mmol/kg gadobutrol (Bayer Pharmaceuticals AG, Berlin, Germany), and two patients received both agents, with 4D flow obtained after ferumoxytol. Ferumoxytol was used as an intravascular contrast agent as clinically indicated. 4D flow imaging data were stored in Digital Imaging and Communications in Medicine (DICOM) format.

2D phase-contrast velocity mapping was performed before contrast administration, with planes prescribed orthogonal to the vessel of interest (echo time [TE] 2.5–4 ms, repetition time [TR] 25–45 ms, field of view 180–300 mm, 30 phases, velocity encoding [VENC] 60–400 cm/s selected by the anticipated peak velocity). Breath-holding was used in non-sedated studies unless patient compliance or clinical status required free-breathing acquisition with signal averaging.

Compressed-sensing 4D flow [23] was acquired after contrast administration in the sagittal projection, with field of view and slice coverage optimized for the clinical indication, three-directional velocity encoding, cross-beam diaphragmatic respiratory navigation, and retrospective electrocardiographic (ECG) gating (TE 2.19–2.41 ms, TR 33.6–51.36 ms, field of view 220–280 mm, voxel size 1.4–1.8 mm, 12–26 phases, flip angle 12–15°, VENC 120–225 cm/s, acceleration factor 4.9–5.7). In two illustrative cases, free-running cardiac- and respiratory-motion-resolved (i.e., 5D) contrast-enhanced whole-heart imaging was also acquired [6] and is shown in the supplementary movies.

**Preprocessing**

4D flow data were imported into SlicerHeart [16] using a purpose-built loader, as detailed under Implementation; the workflow is summarized in Figure 1 and demonstrated in Movies 1 and 2. Each acquisition comprises a magnitude series and three phase series encoding velocity in orthogonal directions. These were grouped and sorted by spatial and temporal position and assembled into a time-resolved series in which each voxel holds a three-component velocity vector at every time point.

Background phase offsets from eddy currents and related effects are corrected by fitting a first- or second-order polynomial to static tissue voxels (those with sufficient magnitude signal and near-zero velocity) and subtracting the fitted field from every frame [24–26]. Phase wraps (aliasing) occur where velocities exceed the VENC value. Because a VENC that permits limited wrap improves use of the dynamic range and the accuracy of low velocities [13], images may be acquired with aliasing in some regions, which must then be corrected in post-processing. After reviewing current unwrapping methods [27], we implemented temporal unwrapping for its simplicity and efficiency [28]. The method detects abrupt jumps in the phase time course of each voxel and stores the result in a label image whose value gives the number of wraps per voxel (Figure 2). Voxels below a user-adjustable magnitude-noise threshold are excluded, and up to three successive wraps in either direction can potentially be recovered. The automated result can be reviewed overlaid on the velocity image and, if needed, edited with standard segmentation tools; corrected images are regenerated within seconds for visual verification.

The polynomial background correction and temporal unwrapping described here were used for all analyses reported in this study. Alternative and additional preprocessing methods, including spatial unwrapping, robust background-correction fits, and divergence-free filtering, are provided as draft implementations in a separate library, described under Implementation and in Appendix B.

**Flow analysis**

Quantitative and visual analysis both begin from a region of interest. The vessel lumen is segmented at a single reference time point, preferably the frame nearest the temporal midpoint of the sequence, which minimizes the registration distance to the remaining frames, using the Segment Editor in 3D Slicer by one of three methods: thresholding, region growing, or an automated deep-learning model [18,29]. The segmentation is propagated to the remaining time points by intensity-based deformable (b-spline) registration to the segmented reference, as detailed under Implementation. For the study cohort, the segmentation method and reference time point were chosen by the analyst for each vessel.

Measurement planes are positioned interactively along vessels or at heart valves. At each plane, points are sampled on a regular grid within a rectangular region defined with the Markups module of 3D Slicer [15] and masked by the vessel segmentation. These sample points are used both to quantify the velocity field and to seed the streamline visualization.

**Visualization**

Volume rendering of the magnitude (tissue) images provides structural context, and the magnitude images, velocity field, segmentation, and measurement planes can be displayed together and animated through the cardiac cycle, building on our prior work [14].

Flow is shown as velocity glyphs and as streamlines [2]. Glyphs (arrows) are displayed on image slices and colored by speed; when a segmentation or speed threshold is applied, the field is masked to the vessel of interest. Streamlines are traced through the 3D velocity field from seeds distributed over the measurement plane, with a configurable maximum length, optional bidirectional integration, and a speed threshold below which tracing stops; they are colored by speed, and their density is set by the seed-point spacing.

A Doppler-like view resamples the velocity field onto a plane and displays the signed through-plane velocity with a cold-to-hot color map analogous to color Doppler ultrasound, so that flow toward and away from the plane is readily distinguished (Figures 3 and 4, Movies 3–5). Absolute speed can be shown instead, using common echocardiographic color scales [30].

**Quantification**

For each measurement plane, hemodynamic metrics are computed at every time point from the velocity samples within the cross-section: cross-sectional area, instantaneous flow rate (separated into forward and backward components), and the mean, minimum, maximum, standard deviation, and selected percentiles of the through-plane velocity. Through-plane velocity, and the forward and backward components derived from it, are defined relative to the orientation of the measurement plane, with flow along the plane normal taken as forward (positive). The contributing region is defined by a speed threshold or, for anatomically specific measurements, by the vessel segmentation.

Time-resolved flow rates are integrated over the cardiac cycle to yield forward, backward, and total stroke volume, regurgitation fraction, and the corresponding volume per minute [3,13], using the heart rate derived from the nominal R-R interval in the DICOM header. Volume per minute is calculated as forward volume multiplied by heart rate, and effective volume per minute as total (net) volume multiplied by heart rate; maximum flow is the peak instantaneous flow rate over the cardiac cycle. Peak velocity and the maximum pressure gradient derived from it by the simplified Bernoulli equation [31] are also reported. All metrics can be displayed as tables or as time-resolved plots.

Because the deformable registration that propagates the segmentation defines a transformation over the whole domain, the same transformation can be applied to the measurement planes, so that each plane follows vessel motion through the cardiac cycle and its cross-section stays aligned with the moving anatomy, analogous to retrospective valve tracking [32].

**Implementation**

The framework is implemented as the “Import 4D Flow” and “Analyze 4D Flow” modules of the open-source SlicerHeart extension for 3D Slicer [15,16], with source code available at www.slicerheart.org. Analyses used 3D Slicer (versions 5.10 and 5.11) with the corresponding SlicerHeart extension releases. In memory, the time-resolved velocity field is represented as a sequence of grid (vector-field) transforms, which 3D Slicer stores and visualizes directly.

These fields, and the deformable transforms that align them, are 5D arrays (three spatial axes, a velocity-vector or displacement component axis, and time). The NRRD reader and writer of ITK [17] previously supported only a single non-spatial axis, so such data required multiple files or application-specific code. As a core contribution, we extended ITK to read and write images, image sequences, vector fields, and transform sequences of five or more dimensions in a single, open, self-describing NRRD file [33]. The functionality is available to any ITK-based tool, including its Python interface, and is used within 3D Slicer (version 5.10 and later) for image- and transform-sequence storage. The complete specification, including the axis-handling strategies and the displacement-field convention, is provided in Appendix A (Sections A.1 to A.8).

Deformable registration across time points uses the SequenceRegistration extension, which performs pairwise b-spline registration of each time point to the reference through SlicerElastix, a 3D Slicer interface to the Elastix toolkit [20].

**Draft implementations of consensus-recommended open processing methods**

Beyond the processing steps described above, we implemented an initial set of preprocessing and quantification methods recommended or discussed in the 4D flow consensus statements [3,13] as slicerheart_4dflow (https://github.com/JolleyLab/slicerheart_4dflow), a standalone, MIT-licensed, open-source Python library that depends only on NumPy and SciPy, is independent of 3D Slicer, and can be called as optional methods by the Import 4D Flow and Analyze 4D Flow modules. The library provides alternative methods for background phase correction, velocity anti-aliasing, noise masking and phase-contrast angiogram generation, divergence-free filtering, and an encoding-direction self-check; through-plane flow quantification with mass-conservation checks; pathline tracing; derived quantities including vorticity, helicity, kinetic energy, viscous energy loss, wall shear stress, and relative pressure; and a per-study quality report. Methods are based on published approaches where available. These methods, their source references, options, defaults, assumptions, and limitations are summarized in Appendix B and documented in the repository.

The library is tested by an automated test suite with two tiers. The synthetic tier comprises analytic phantoms with known pressure, wall shear stress, flow split, aliasing, and background offset [34–36], a phase-contrast forward model that generates velocity noise, aliasing, and partial-volume effects from the signal model [37], and an aortic CFD ground truth [38]; it runs on every change to the repository, and every recorded metric is compared with a stored baseline so that regressions are detected and new methods can be compared with existing ones on the same cases. The clinical tier runs the library methods on measured 4D flow CMR studies exported from the SlicerHeart modules, for which no ground truth exists, and checks internal consistency (residual static-tissue velocity after correction, agreement of the unwrapping with the reviewed wrap counts, VENC adequacy, inflow-outflow balance and flow ratios) and agreement of plane flow with the SlicerHeart modules and, where available, with reference measurements from other software. Both tiers currently pass in full, the clinical tier on 15 compressed sensing 4D flow studies acquired at our institution on a Siemens scanner (the 13 quantified here plus two additional). GE and Philips data will be added as they become available. Because synthetic data cannot represent vendor-specific conventions or scanner noise, and clinical data carry no ground truth, these methods are released as draft implementations to be inspected, compared, and refined by the community rather than as validated tools; their performance across vendors, sequences, and patient populations has not been formally established. The results reported in this study were obtained with the module implementations of temporal unwrapping, polynomial background phase correction, and plane-based quantification described above; the Appendix B methods were not applied to the data used for the comparison with cvi42 and 2D phase contrast.

### Export to inform computational modeling

Segmentations, time-resolved velocity fields, and measurement-plane geometry can be exported in open formats for external analysis and simulation. Vessel-lumen segmentations are exported as surface meshes (e.g., stereolithography [STL] or Visualization Toolkit [VTK] format), the time-resolved velocity field as a 5D NRRD file (.seq.nrrd), and each measurement plane's position, orientation, and size as a JavaScript Object Notation (JSON) file, with time-resolved through-plane velocity and other flow metrics as comma-separated values (CSV) files. These can provide patient-specific geometry, inlet and outlet velocity profiles, and reference velocity fields for open cardiovascular simulation platforms such as SimVascular [39] and FEBio [11,40]. Because the deformable registration defines a transformation over the whole imaged domain, exported measurement planes retain their motion across the cardiac cycle and can support moving-boundary conditions (Figure 5).

**Comparison with commercial software**

Flow quantification from the framework was compared with cvi42 (Circle Cardiovascular Imaging, Calgary, Canada), a widely used commercial package. Framework measurements used the preprocessing and quantification pipeline described above, without the draft methods of Appendix B. In cvi42 (4D Flow module, version 6.0.3), anti-aliasing and static background correction were enabled, with vessels and spatial wrap excluded from the static-tissue mask; the vasculature was thresholded, planes orthogonal to each vessel were prescribed in double-oblique fashion, and regions of interest were contoured across all phases. Analyses on both platforms were performed by four research technicians, and all segmentations and plane placements were reviewed for accuracy by two pediatric cardiologists specializing in CMR. To maximize comparability, measurement planes were placed at visually matched locations on both platforms (Figure 6), so analysts were not blinded between platforms. 2D phase-contrast measurements were taken from clinical reports generated in routine care with the cvi42 2D Flow module.

For each patient, measurements were matched by anatomical label across aortic, pulmonary arterial, systemic venous, and pulmonary venous segments from the exported reports of each platform, and all regions of interest with measurements available from both the framework and cvi42 4D flow were included, with none excluded on the basis of agreement. The primary comparison was the framework versus cvi42 4D flow, with forward volume and effective volume per minute as primary endpoints and mean through-plane velocity, maximum flow, and cross-sectional area as additional metrics (Table 2); backward, total, net positive, and net negative volume, regurgitation fraction, volume per minute, peak velocity, and maximum pressure gradient are reported as secondary analyses (Supplementary Table S1). Forward, backward, and total volume, regurgitation fraction, effective volume per minute, and peak velocity were also compared with 2D phase contrast for both the framework and cvi42 4D flow in the subset of included regions with paired 2D measurements, and heart rate was compared between cvi42 4D flow and 2D phase contrast (Supplementary Table S2).

**Statistical analysis**

Each region of interest was measured once on each platform; regions without paired framework and cvi42 4D flow values were not included. For every paired measurement, the difference was taken as the framework value minus the cvi42 value. Time-resolved metrics (cross-sectional area and mean velocity) were averaged across cardiac phases matched by trigger time between platforms. Agreement for each metric was summarized across all regions by Bland-Altman analysis, reporting the mean difference (bias) and 95% limits of agreement, together with the mean absolute percentage error relative to the cvi42 value. For volume-derived secondary metrics (backward, total, and net volume), percentage error was expressed relative to the cvi42 forward volume, so that related quantities shared a common denominator and percentages remained stable when a component approached zero. Regurgitation fraction, already a percentage, was compared as a difference in percentage points, with the cvi42 value taken as 0% when cvi42 reported no backward flow. Agreement was additionally quantified with the intraclass correlation coefficient (ICC; two-way mixed effects, absolute agreement, single measurement) [41]. Because multiple regions were analyzed per patient, observations were not independent. Limits of agreement were therefore adjusted for within-patient clustering using a random-effects variance-components approach adapted from the method of Bland and Altman for multiple observations per individual, in the form for a true value that varies between observations, since each region represents a different vessel [42], and 95% confidence intervals (CIs) for the bias, limits of agreement, and ICC were estimated by patient-level cluster bootstrap (20,000 resamples). Linear regression of cvi42 against framework values, with the coefficient of determination ($R^2$), was used to visualize proportional bias relative to the line of identity, as a descriptive complement to the Bland-Altman analysis (Figure 7). Agreement with 2D phase contrast was assessed in the same way for both the framework and cvi42 4D flow, with 2D phase contrast as the reference. No hypothesis tests were performed, and no formal sample size calculation was undertaken for this technical development study. Analyses were performed with custom Python scripts (Python 3.12.10 with NumPy and openpyxl; ICC and bootstrap calculations were implemented directly), and figures were generated with Matplotlib.

## Results

### A unified open-source framework for time-resolved (4D) CMR

Our work resulted in an open-source framework for the import, visualization, and quantification of time-resolved HDim CMR, released as the “Import 4D Flow” and “Analyze 4D Flow” modules in the SlicerHeart extension for 3D Slicer (Movies 1 and 2), with source code available at www.slicerheart.org. In addition, we have released the slicerheart_4dflow Python package to support these workflows (Appendix B). The SlicerHeart framework brings the full post-processing workflow into a single graphical environment, from reconstruction of time-resolved data from DICOM through automated preprocessing, interactive visualization, and quantification of both cardiac structure and flow. Across the cohort, each 4D flow study was imported as a time-resolved velocity field, stored as a 5D array, and carried through the complete workflow without transfer of data between separate applications (Figure 1). Segmentation, deformable registration, volume rendering, and quantification operated on shared data structures, so that tissue and flow were examined, measured, and modeled together within one environment.

### Native HDim array support in ITK

As a core infrastructure contribution, we extended the NRRD reader and writer of ITK to natively read and write images, image sequences, vector fields, and transform sequences in 4D flow and HDim CMR in a single, self-describing file. Time-resolved image, velocity-vector, and displacement-field sequences, comprising three spatial axes, a component axis, and time, can now be stored and interchanged in one open format rather than as multiple files or through application-specific code. The implementation preserves the role, units, and metadata of every axis and tags displacement fields for consistent interpretation. The functionality is available in ITK, including its Python interface, and is used within 3D Slicer (version 5.10 and later) for image- and transform-sequence storage, so that any ITK-based tool can read and write these data without custom development. The complete specification is provided in Appendix A (Sections A.1 to A.8).

### Demonstration across congenital heart disease

We demonstrate the utility of the SlicerHeart modules applied to 4D flow and 5D CMR across a range of congenital lesions (Figures 3 and 4, Movies 3–5). Temporal phase unwrapping identified and labeled aliased voxels for review and correction, recovering physiologic velocity fields in regions of wrap (Figure 2). Volume rendering of the magnitude images provided structural context, and the magnitude images, velocity field, segmentation, and measurement planes were displayed together and animated through the cardiac cycle (Figure 1). Flow was shown as velocity glyphs, as streamlines seeded from the measurement plane at configurable density, and as a Doppler-like through-plane rendering that distinguished flow toward and away from the plane (Figures 2–4). In a patient with repaired truncus arteriosus, simultaneous rendering of tissue and flow depicted diastolic regurgitation through a quadricuspid truncal valve (Figure 3, Movie 3), and in a patient with unbalanced atrioventricular canal defect, systolic atrioventricular valve regurgitation (Movie 4). In a patient with hypoplastic left heart syndrome, the framework depicted flow across a stented atrial septal communication within the rendered anatomy (Figure 4, Movie 5). The framework can also export time-resolved anatomy, velocity fields, segmentations, and measurement-plane geometry in open formats suitable as patient-specific geometry, velocity profiles, and reference hemodynamics for simulation platforms (Figure 5).

**Validation: agreement with commercial software and 2D phase contrast**

Thirteen patients (6 male, 7 female) were included (Table 1). Diagnoses spanned isolated septal defects through single-ventricle physiology and included atrial and ventricular septal defects, double outlet right ventricle, Ebstein anomaly, pulmonary atresia with intact ventricular septum, total anomalous pulmonary venous connection, congenitally corrected transposition of the great arteries, and heterotaxy with single-ventricle physiology palliated toward a Fontan circulation. Age ranged from 0.04 to 17.1 years (median 3.9 years, interquartile range [IQR] 0.4 to 5.6), and five patients were infants under one year of age. Ten of the 13 patients had undergone prior surgical or catheter-based intervention.

The same vessels and visually comparable plane locations were analyzed in the framework and in cvi42 (version 6.0.3). A total of 108 regions of interest with paired framework and cvi42 4D flow measurements were analyzed across aortic, pulmonary arterial, systemic venous, and pulmonary venous segments, with a median of 8 and mean of 8.3 regions per patient (Figure 6, Table 2). Of these, 65 were arterial (29 aortic, including 7 at the aortic valve, and 36 pulmonary arterial) and 43 were venous (27 systemic venous and 16 pulmonary venous).

Flow measurements from the framework agreed closely with cvi42 for the primary endpoints, with small biases whose 95% CIs included zero. Forward volume showed a bias of −0.1 mL (95% CI −0.3 to 0.1; 95% limits of agreement 2.2 to −2.3), and effective volume per minute a bias of −0.015 L/min (−0.038 to 0.004; 0.21 to −0.24), each small relative to the measured range. Maximum flow likewise showed a small bias with a 95% CI that included zero (1.2 mL/s; −0.5 to 2.8; 16.8 to −14.4). Small biases with 95% CIs excluding zero were observed for cross-sectional area, which was larger with the framework (bias 5.1 $mm^2$; 1.5 to 8.5; 46.5 to −36.3), and mean velocity, which was lower (−2.0 cm/s; −4.8 to −0.6; 9.8 to −13.9). Mean absolute percentage error was 4.8% for forward volume, 6.4% for effective volume per minute, 9.2% for mean velocity, 9.0% for maximum flow, and 11.1% for cross-sectional area. The ICCs were 0.997 (95% CI 0.990 to 0.999) for forward volume, 0.994 (0.988 to 0.997) for effective volume per minute, 0.795 (0.560 to 0.973) for mean velocity, 0.993 (0.981 to 0.998) for maximum flow, and 0.986 (0.972 to 0.991) for cross-sectional area (Table 2). Forward volume agreement was similar for arterial and venous regions, with slightly higher percentage error in venous regions (bias −0.3 and 0.3 mL; mean absolute percentage error 4.3% and 5.7%; ICC 0.998 and 0.993, respectively). On Bland-Altman analysis, differences were distributed around the bias line without a consistent trend across the measured range for the volumetric flow metrics, whereas maximum flow showed greater scatter at higher values (Figure 7). The identity-line regression was consistent with these findings (Figure 7).

Among secondary metrics, agreement was close for backward, total, and net positive volume and for volume per minute (ICC 0.950 to 0.996), and for regurgitation fraction (ICC 0.915; mean absolute difference 2.5 percentage points). Agreement was weaker for velocity-derived and small-volume metrics: peak velocity was lower with the framework (bias −10.0 cm/s; 95% CI −15.7 to −5.9; ICC 0.836), with corresponding differences in maximum pressure gradient (ICC 0.765), and net negative volume showed an ICC of 0.811 (Supplementary Table S1).

Forward volume was further compared with 2D phase contrast in the 100 regions of interest, from all 13 patients that had paired 2D measurements. The mean difference in heart rate between 4D flow (as reported by cvi42) and 2D phase contrast was 4.1 beats per minute (bpm; 95% CI −2.9 to 13.3), although differences between individual acquisitions were wider (95% limits of agreement 33.5 to −25.3 bpm; ICC 0.871; Supplementary Table S2). Relative to 2D phase contrast, the framework showed a bias of −2.2 mL (95% CI −3.2 to −1.4; 95% limits of agreement 2.7 to −7.2), a mean absolute percentage error of 21.8%, and an ICC of 0.978 (0.934 to 0.991). In the same regions of interest, cvi42 4D flow showed a bias of −2.1 mL (−3.0 to −1.3; 2.6 to −6.8), a mean absolute percentage error of 21.3%, and an ICC of 0.980 (0.940 to 0.992). Both 4D methods therefore showed similar bias, percentage error, and limits of agreement relative to 2D phase contrast. Agreement with 2D phase contrast for the remaining metrics is reported in Supplementary Table S2.

## Discussion

We present an open-source framework for the import, visualization, and quantification of time-resolved HDim CMR, encompassing both anatomic (tissue) imaging and 4D flow, implemented within the SlicerHeart extension for 3D Slicer. It provides a complete post-processing pipeline: reconstruction of time-resolved data from DICOM, phase unwrapping and background-phase correction of velocity data, vessel segmentation with registration-based propagation across the cardiac cycle, and interactive visualization and quantification of structure and flow. In a pediatric cohort spanning a broad range of CHD, flow measurements from the framework agreed closely with those of a widely used commercial package across the primary endpoints. Underlying this work, we extended ITK to natively represent five- and higher-dimensional arrays, so that time-resolved images, vector fields, and transforms share one open format. Finally, a companion library provides open implementations of common post-processing methods with internal-consistency checks (Appendix B), an unmet need identified in the 2023 consensus statement [3].

Post-processing of 4D flow CMR is currently dominated by commercial applications that provide validated, clinically oriented quantification and agree well with 2D phase-contrast reference measurements in patients with CHD [43]. These packages are, however, typically closed source and standalone, so they cannot be readily inspected, extended, or coupled to other analyses by the research community, and their results can differ meaningfully from one another, limiting comparability and reproducibility [12]. They are also dedicated to flow and are not designed for exploring how best to display and quantify the time-resolved anatomic datasets produced by emerging acquisitions such as multiphase steady-state imaging with contrast enhancement (MUSIC) and free-running, motion-resolved whole-heart imaging. Open-source efforts have partly addressed accessibility: an early open tool provided 4D flow quantification but depended in part on proprietary components [44]. To our knowledge, however, no current platform situates time-resolved anatomic and flow analysis together within a general-purpose image-computing platform in which segmentation, registration, visualization, and modeling interoperate.

Our framework addresses this need. It unifies the workflow from import and automated preprocessing through interactive visualization to quantification in a single graphical application, so that researchers can move from raw acquisition to structural and hemodynamic analysis without leaving one environment. In doing so, it matures the volume-rendering and flow visualizations we prototyped previously [14] into a robust, user-friendly, and extensible tool, and adds quantification and registration-based measurement. Because it is built within 3D Slicer and SlicerHeart [15,16], it inherits mature segmentation, deformable registration, volume rendering, and cardiac-modeling capabilities, so that tissue and flow are examined, measured, and modeled together rather than in isolation. For example, a single segmentation, propagated across the cardiac cycle by deformable registration, drives both measurement-plane tracking and quantification, and visualizations such as dense streamlines or a color Doppler-like velocity rendering can be reconfigured for specific anatomic questions.

Agreement of our initial framework with commercial software was closest for the primary volumetric endpoints, for which bias was small, with 95% CIs including zero, and weaker for velocity-based metrics. The framework measured slightly larger cross-sectional areas and lower mean velocities, a pattern consistent with a marginally more inclusive lumen contour that incorporates slower near-wall flow; because flow is the product of area and velocity, these offsets largely cancel, and forward volume and maximum flow showed small biases with 95% CIs including zero. Peak velocity, which depends on a single voxel, and the pressure gradient derived from it were most sensitive to differences in segmentation and plane position. Forward volume percentage differences from 2D phase contrast were similar for both 4D methods (21.8% and 21.3%) and larger than those between them (4.8%), consistent with differences between the 4D and 2D measurements, including timing, heart rate, respiratory compensation, the number of heartbeats sampled, spatial and temporal resolution, and independent plane placement, rather than with the 4D post-processing software.

The HDim array support in ITK is an infrastructure contribution in its own right. Time-resolved velocity fields, and the deformable transforms that align them, are inherently high-dimensional, yet ITK's NRRD reader and writer previously supported only a single non-spatial axis, forcing such data into multiple files or application-specific code. We extended them to read and write images, vector fields, and transform sequences of five or more dimensions in a single, self-describing file, preserving the role, units, and metadata of every axis and tagging displacement fields for consistent interpretation. Because this capability resides in ITK rather than in our application, any ITK-based tool can use it, and any time-resolved vector or tensor field is now representable in one open format. Our modules build on this foundation, exposing it to users in interactive, graphical form within an open image-computing platform.

A further advantage is transparency and extensibility. Because the source code and processing algorithms are fully open, each step, from phase unwrapping and background correction to plane-based quantification, can be inspected, modified, and improved by the research community. This supports development of new processing algorithms and adaptation to acquisitions from vendors beyond the single-vendor data used here, and positions the framework as a common, inspectable substrate for harmonizing time-resolved CMR post-processing across centers, a recognized source of variability in the field [12]. The same architecture invites community extension to advanced derived parameters, such as wall shear stress, kinetic energy, and viscous energy or power loss, initial draft implementations of which are provided in Appendix B. These hemodynamic markers are of growing interest in CHD and have been linked to outcomes in single-ventricle physiology, repaired tetralogy of Fallot, and bicuspid aortic valve disease [7,9,45–49].

A distinguishing capability is the framework's suitability for informing and validating computational simulation. Patient-specific CFD and FSI models are increasingly used to study hemodynamics and to guide surgical and interventional planning in CHD, from single-ventricle physiology to valvular and aortic disease [8,11], yet their reliability depends on validation against in vivo data and on non-invasive, patient-specific boundary conditions. CMR-derived anatomy and flow have been used to prescribe these conditions in patients [9] and to verify simulation output against velocity fields measured in vitro [10]. By exporting time-resolved anatomy, velocity fields, segmentations, and measurement-plane geometry in open formats, the framework can supply patient-specific geometry, velocity profiles, and reference hemodynamics to open simulation platforms [11,39], and its moving measurement planes support moving-boundary conditions. Bringing imaging, segmentation, and simulation into a single environment lowers the barrier to closed-loop comparison between measured and computed flow, complementing ongoing multi-physics modeling of valves, devices, and myocardium in SlicerHeart as a step toward patient-specific digital twins in CHD [11,50–52].

Looking further ahead, the same open, interoperable design positions the framework to interface with emerging machine-learning workflows. The 3D Slicer ecosystem already supports deep-learning segmentation [18], and the ability to export co-registered anatomy, segmentations, and time-resolved velocity fields in open formats provides ready-made, labeled inputs for model development. One promising direction is the training of neural surrogates that learn the mapping from anatomy and boundary conditions to hemodynamics by combining physically grounded simulation with in vivo measurement [53–55]. Physics-informed neural networks, which infer hidden velocity and pressure fields from sparse flow data [56], have already been applied to 4D flow CMR to estimate pressure, calibrate simulation boundary conditions, and enhance resolution [57,58], and large-scale foundation and surrogate models trained on heterogeneous physical data now approach or exceed conventional numerical solvers at a fraction of their computational cost [59,60]. The framework can thus supply the paired clinical and simulated data needed to train and validate such surrogates, a step toward near-real-time, patient-specific hemodynamic prediction in congenital and structural heart disease.

### Limitations

This work is a technical development validated in a small, single-center pediatric cohort acquired on a single vendor's scanners; the open design is intended to enable the multi-vendor, multi-center evaluation needed to establish generalizability. Comparison was performed against one commercial package, which is itself closed source, so agreement was assessed relative to a practical reference rather than an absolute ground truth, and validation against known flow rates in physical phantoms is needed. Measurement planes were deliberately placed at visually matched locations on both platforms, and analysts were not blinded between platforms; results therefore reflect agreement under matched conditions, and inter- and intraobserver reproducibility were not assessed. 2D phase-contrast values were taken from clinical reports, acquired at different times and heart rates and analyzed independently of the 4D flow data. Multiple regions of interest were analyzed per patient; although limits of agreement and CIs accounted for this clustering, agreement by vessel type is reported descriptively and was not formally compared, given the limited number of patients. Accuracy depends on acquisition quality and on upstream steps such as phase unwrapping and background-phase correction, which, although automated, may require manual review, and registration-based propagation of segmentations can degrade in regions of low signal or highly disturbed flow. Several additional methods are provided only as draft implementations (Appendix B) and were neither applied to nor validated in this study. The framework is a research tool and is not cleared for clinical use. Finally, the CFD, FSI, and machine-learning applications are presented as enabling capabilities rather than validated in prospective studies, which are the subject of ongoing work.

## Conclusions

This framework brings the import, visualization, and quantification of time-resolved 4D flow and other HDim CMR, spanning both anatomy and flow, into the open 3D Slicer and SlicerHeart ecosystem, enabled by native high-dimensional array support contributed to ITK, and agrees closely with a widely used commercial package. By making analysis transparent, interoperable with segmentation, modeling, and simulation tools, and freely available to any center, it provides a foundation for reproducible, harmonizable analysis and for the validation of computational models, with particular relevance to the complex, heterogeneous anatomy of congenital and structural heart disease.

## Funding

NIH 1R01HL153166, a Children's Hospital of Philadelphia Post-Frontier Award, the Cora Topolewski Fund at the CHOP Pediatric Valve Center, the Topolewski Endowed Chair in Pediatric Cardiology, and The Mussallem Congenital Heart Disease Alliance.

## Declaration of competing interest

Csaba Pinter is an employee of Ebatinca S.L. The other authors declare that they have no known competing financial interests or personal relationships that could have appeared to influence the work reported in this paper.

## Ethics approval and consent to participate

The human subjects research data used in this study were collected under protocol IRB-16-012853, reviewed by the Committees for the Protection of Human Subjects (Institutional Review Board) of the Children's Hospital of Philadelphia on July 8, 2016. The study was determined to meet exemption criteria under 45 CFR 46.104(d)(4)(iii); therefore, consent and assent were not required. A waiver of Health Insurance Portability and Accountability Act (HIPAA) authorization was granted under 45 CFR 164.512(i)(2)(ii) for access to identifiable information from medical records. The study was conducted in accordance with the Declaration of Helsinki. No individually identifiable patient information is included in the manuscript or supplementary materials.

## Data and code availability

The framework is open-source and freely available as part of the SlicerHeart extension (www.slicerheart.org) for 3D Slicer (www.slicer.org). The five-dimensional (5D) array support is available in the Insight Toolkit (ITK) and in 3D Slicer (version 5.10 and later). The draft filters and testing suite with synthetic phantoms and imaging data is available at https://github.com/JolleyLab/slicerheart_4dflow.

## Credit authorship contribution statement

Andras Lasso: Conceptualization, Methodology, Software, Validation, Visualization, Writing – original draft, Writing – review & editing. Julia Iacovella: Conceptualization, Methodology, Investigation, Formal analysis, Data curation, Visualization, Writing – original draft, Writing – review & editing. Csaba Pinter: Software, Methodology, Writing – review & editing. Kevin K. Whitehead: Resources, Investigation, Validation, Data curation, Writing – review & editing. Sehdev Pressley: Investigation, Data curation, Writing – review & editing. Danish Vaiyani: Resources, Investigation, Validation, Data curation, Writing – review & editing. Matthew Daemer: Investigation, Data curation, Writing – review & editing. Katherine Hurm: Investigation, Data curation, Writing – review & editing. Analise M. Sulentic: Investigation, Data curation, Writing – review & editing. Christian Herz: Software, Methodology, Writing – review & editing. Ryan Black: Methodology, Visualization, Writing – review & editing. Seda Aslan: Methodology, Visualization, Writing – review & editing. Chris Roy: Methodology, Resources, Writing – review & editing. Mark A. Fogel: Methodology, Writing – review & editing. Jing Huang: Methodology, Formal analysis, Writing – review & editing. Matthew A. Jolley: Conceptualization, Methodology, Supervision, Project administration, Funding acquisition, Resources, Writing – original draft, Writing – review & editing. Andras Lasso and Julia Iacovella contributed equally to this work. All authors read and approved of the final manuscript.

## Acknowledgements

We appreciate the contributions of 3D Slicer, SlicerHeart, and the VTK and ITK open-source communities.

## Declaration of generative AI and AI-assisted technologies in the manuscript preparation process

During the preparation of this work, the authors used Claude Opus 5.5 and Claude Fable 5.1 (Anthropic) to assist with editing text (including figure and table legends), identifying typographical errors and structural deficiencies, checking internal consistency of reported values, and managing references; these tools were also used to summarize source code for the appendices. GitHub Copilot in Visual Studio Code (agent mode, GPT-6 Luna) and Claude (Anthropic) were used to assist with writing and running the statistical analysis and comparison scripts and generating tables of results. After using these tools, the authors reviewed and edited the content as needed and take full responsibility for the content of the published article.

## References

[1] Fratz S, Chung T, Greil GF, Samyn MM, Taylor AM, Valsangiacomo Buechel ER, et al. Guidelines and protocols for cardiovascular magnetic resonance in children and adults with congenital heart disease: SCMR expert consensus group on congenital heart disease. J Cardiovasc Magn Reson Off J Soc Cardiovasc Magn Reson 2013;15:51. https://doi.org/10.1186/1532-429X-15-51.

[2] Markl M, Frydrychowicz A, Kozerke S, Hope M, Wieben O. 4D flow MRI. J Magn Reson Imaging JMRI 2012;36:1015–36. https://doi.org/10.1002/jmri.23632.

[3] Bissell MM, Raimondi F, Ait Ali L, Allen BD, Barker AJ, Bolger A, et al. 4D Flow cardiovascular magnetic resonance consensus statement: 2023 update. J Cardiovasc Magn Reson Off J Soc Cardiovasc Magn Reson 2023;25:40. https://doi.org/10.1186/s12968-023-00942-z.

[4] Han F, Rapacchi S, Khan S, Ayad I, Salusky I, Gabriel S, et al. Four-dimensional, multiphase, steady-state imaging with contrast enhancement (MUSIC) in the heart: a feasibility study in children. Magn Reson Med 2015;74:1042–9. https://doi.org/10.1002/mrm.25491.

[5] Nguyen K-L, Ghosh RM, Griffin LM, Yoshida T, Bedayat A, Rigsby CK, et al. Four-dimensional Multiphase Steady-State MRI with Ferumoxytol Enhancement: Early Multicenter Feasibility in Pediatric Congenital Heart Disease. Radiology 2021;300:162–73. https://doi.org/10.1148/radiol.2021203696.

[6] Roy CW, Di Sopra L, Whitehead KK, Piccini D, Yerly J, Heerfordt J, et al. Free-running cardiac and respiratory motion-resolved 5D whole-heart coronary cardiovascular magnetic resonance angiography in pediatric cardiac patients using ferumoxytol. J Cardiovasc Magn Reson Off J Soc Cardiovasc Magn Reson 2022;24:39. https://doi.org/10.1186/s12968-022-00871-3.

[7] Whitehead KK, Pekkan K, Kitajima HD, Paridon SM, Yoganathan AP, Fogel MA. Nonlinear power loss during exercise in single-ventricle patients after the Fontan: insights from computational fluid dynamics. Circulation 2007;116:I165-171. https://doi.org/10.1161/CIRCULATIONAHA.106.680827.

[8] Marsden AL, Bernstein AJ, Reddy VM, Shadden SC, Spilker RL, Chan FP, et al. Evaluation of a novel Y-shaped extracardiac Fontan baffle using computational fluid dynamics. J Thorac Cardiovasc Surg 2009;137:394-403.e2. https://doi.org/10.1016/j.jtcvs.2008.06.043.

[9] Haggerty CM, Restrepo M, Tang E, de Zélicourt DA, Sundareswaran KS, Mirabella L, et al. Fontan hemodynamics from 100 patient-specific cardiac magnetic resonance studies: a computational fluid dynamics analysis. J Thorac Cardiovasc Surg 2014;148:1481–9. https://doi.org/10.1016/j.jtcvs.2013.11.060.

[10] Kaiser AD, Schiavone NK, Elkins CJ, McElhinney DB, Eaton JK, Marsden AL. Comparison of Immersed Boundary Simulations of Heart Valve Hemodynamics Against In Vitro 4D Flow MRI Data. Ann Biomed Eng 2023;51:2267–88. https://doi.org/10.1007/s10439-023-03266-2.

[11] Black RT, Maas SA, Wu W, Maheshwari J, Kolev T, Weiss JA, et al. An open-source computational framework for immersed fluid-structure interaction modeling using FEBio and MFEM. Eng Comput 2026;42:166. https://doi.org/10.1007/s00366-026-02371-9.

[12] Oechtering TH, Nowak A, Sieren MM, Stroth AM, Kirschke N, Wegner F, et al. Repeatability and reproducibility of various 4D Flow MRI postprocessing software programs in a multi-software and multi-vendor cross-over comparison study. J Cardiovasc Magn Reson Off J Soc Cardiovasc Magn Reson 2023;25:22. https://doi.org/10.1186/s12968-023-00921-4.
[13] Dyverfeldt P, Bissell M, Barker AJ, Bolger AF, Carlhäll C-J, Ebbers T, et al. 4D flow cardiovascular magnetic resonance consensus statement. J Cardiovasc Magn Reson Off J Soc Cardiovasc Magn Reson 2015;17:72. https://doi.org/10.1186/s12968-015-0174-5.
[14] Iacovella J, Vaiyani D, Pressley S, Lasso A, Sulentic AM, Cianciulli AR, et al. Rapid Visualization of Valves and Myocardium Using Volume Rendering of 3D Cardiac MRI, 4D Cine, and 4D Flow Images. Radiol Cardiothorac Imaging 2026;8:e250129. https://doi.org/10.1148/ryct.250129.
[15] Fedorov A, Beichel R, Kalpathy-Cramer J, Finet J, Fillion-Robin J-C, Pujol S, et al. 3D Slicer as an image computing platform for the Quantitative Imaging Network. Magn Reson Imaging 2012;30:1323–41. https://doi.org/10.1016/j.mri.2012.05.001.
[16] Lasso A, Herz C, Nam H, Cianciulli A, Pieper S, Drouin S, et al. SlicerHeart: An open-source computing platform for cardiac image analysis and modeling. Front Cardiovasc Med 2022;9:886549. https://doi.org/10.3389/fcvm.2022.886549.
[17] McCormick M, Liu X, Jomier J, Marion C, Ibanez L. ITK: enabling reproducible research and open science. Front Neuroinformatics 2014;8:13. https://doi.org/10.3389/fninf.2014.00013.
[18] Diaz-Pinto A, Alle S, Nath V, Tang Y, Ihsani A, Asad M, et al. MONAI Label: A framework for AI-assisted interactive labeling of 3D medical images. Med Image Anal 2024;95:103207. https://doi.org/10.1016/j.media.2024.103207.
[19] Wasserthal J, Breit H-C, Meyer MT, Pradella M, Hinck D, Sauter AW, et al. TotalSegmentator: Robust Segmentation of 104 Anatomic Structures in CT Images. Radiol Artif Intell 2023;5:e230024. https://doi.org/10.1148/ryai.230024.
[20] Klein S, Staring M, Murphy K, Viergever MA, Pluim JPW. elastix: a toolbox for intensity-based medical image registration. IEEE Trans Med Imaging 2010;29:196–205. https://doi.org/10.1109/TMI.2009.2035616.
[21] Avants BB, Tustison NJ, Song G, Cook PA, Klein A, Gee JC. A reproducible evaluation of ANTs similarity metric performance in brain image registration. NeuroImage 2011;54:2033–44. https://doi.org/10.1016/j.neuroimage.2010.09.025.
[22] Pinter C, Lasso A, Choueib S, Asselin M, Fillion-Robin J-C, Vimort J-B, et al. SlicerVR for Medical Intervention Training and Planning in Immersive Virtual Reality. IEEE Trans Med Robot Bionics 2020;2:108–17. https://doi.org/10.1109/tmrb.2020.2983199.
[23] Ma LE, Markl M, Chow K, Huh H, Forman C, Vali A, et al. Aortic 4D flow MRI in 2 minutes using compressed sensing, respiratory controlled adaptive k-space reordering, and inline reconstruction. Magn Reson Med 2019;81:3675–90. https://doi.org/10.1002/mrm.27684.
[24] Walker PG, Cranney GB, Scheidegger MB, Waseleski G, Pohost GM, Yoganathan AP. Semiautomated method for noise reduction and background phase error correction in MR phase velocity data. J Magn Reson Imaging JMRI 1993;3:521–30. https://doi.org/10.1002/jmri.1880030315.
[25] Lankhaar J-W, Hofman MBM, Marcus JT, Zwanenburg JJM, Faes TJC, Vonk-Noordegraaf A. Correction of phase offset errors in main pulmonary artery flow quantification. J Magn Reson Imaging JMRI 2005;22:73–9. https://doi.org/10.1002/jmri.20361.
[26] Pruitt AA, Jin N, Liu Y, Simonetti OP, Ahmad R. A method to correct background phase offset for phase-contrast MRI in the presence of steady flow and spatial wrap-around artifact. Magn Reson Med 2019;81:2424–38. https://doi.org/10.1002/mrm.27572.

[27] Löcke M, Garay Labra JE, Franco P, Uribe S, Bertoglio C. A comparison of phase unwrapping methods in velocity-encoded MRI for aortic flows. Magn Reson Med 2023;90:2102–15. https://doi.org/10.1002/mrm.29767.

[28] Xiang QS. Temporal phase unwrapping for CINE velocity imaging. J Magn Reson Imaging JMRI 1995;5:529–34. https://doi.org/10.1002/jmri.1880050509.

[29] Akinci D'Antonoli T, Berger LK, Indrakanti AK, Vishwanathan N, Weiss J, Jung M, et al. TotalSegmentator MRI: Robust Sequence-independent Segmentation of Multiple Anatomic Structures in MRI. Radiology 2025;314:e241613. https://doi.org/10.1148/radiol.241613.

[30] Mitchell C, Rahko PS, Blauwet LA, Canaday B, Finstuen JA, Foster MC, et al. Guidelines for Performing a Comprehensive Transthoracic Echocardiographic Examination in Adults: Recommendations from the American Society of Echocardiography. J Am Soc Echocardiogr Off Publ Am Soc Echocardiogr 2019;32:1–64. https://doi.org/10.1016/j.echo.2018.06.004.

[31] Hatle L, Brubakk A, Tromsdal A, Angelsen B. Noninvasive assessment of pressure drop in mitral stenosis by Doppler ultrasound. Br Heart J 1978;40:131–40. https://doi.org/10.1136/hrt.40.2.131.

[32] Westenberg JJM, Roes SD, Ajmone Marsan N, Binnendijk NMJ, Doornbos J, Bax JJ, et al. Mitral valve and tricuspid valve blood flow: accurate quantification with 3D velocity-encoded MR imaging with retrospective valve tracking. Radiology 2008;249:792–800. https://doi.org/10.1148/radiol.2492080146.

[33] Kindlmann G. NRRD: Definition of NRRD File Format. Teem n.d. https://teem.sourceforge.net/nrrd/format.html (accessed October 23, 2026).

[34] Womersley JR. Method for the calculation of velocity, rate of flow and viscous drag in arteries when the pressure gradient is known. J Physiol 1955;127:553–63. https://doi.org/10.1113/jphysiol.1955.sp005276.

[35] Young DF, Tsai FY. Flow characteristics in models of arterial stenoses. I. Steady flow. J Biomech 1973;6:395–410. https://doi.org/10.1016/0021-9290(73)90099-7.

[36] Murray CD. The Physiological Principle of Minimum Work: I. The Vascular System and the Cost of Blood Volume. Proc Natl Acad Sci U S A 1926;12:207–14. https://doi.org/10.1073/pnas.12.3.207.

[37] Pelc NJ, Bernstein MA, Shimakawa A, Glover GH. Encoding strategies for three-direction phase-contrast MR imaging of flow. J Magn Reson Imaging JMRI 1991;1:405–13. https://doi.org/10.1002/jmri.1880010404.

[38] Ferdian E, Suinesiaputra A, Dubowitz DJ, Zhao D, Wang A, Cowan B, et al. 4DFlowNet: Super-Resolution 4D Flow MRI Using Deep Learning and Computational Fluid Dynamics. Front Phys 2020;8:138. https://doi.org/10.3389/fphy.2020.00138.

[39] Updegrove A, Wilson NM, Merkow J, Lan H, Marsden AL, Shadden SC. SimVascular: An Open Source Pipeline for Cardiovascular Simulation. Ann Biomed Eng 2017;45:525–41. https://doi.org/10.1007/s10439-016-1762-8.

[40] Maas SA, Ellis BJ, Ateshian GA, Weiss JA. FEBio: finite elements for biomechanics. J Biomech Eng 2012;134:011005. https://doi.org/10.1115/1.4005694.

[41] Koo TK, Li MY. A Guideline of Selecting and Reporting Intraclass Correlation Coefficients for Reliability Research. J Chiropr Med 2016;15:155–63. https://doi.org/10.1016/j.jcm.2016.02.012.

[42] Bland JM, Altman DG. Agreement Between Methods of Measurement with Multiple Observations Per Individual. J Biopharm Stat 2007;17:571–82. https://doi.org/10.1080/10543400701329422.

[43] Burkhardt BEU, Kellenberger CJ, Callaghan FM, Valsangiacomo Buechel ER, Geiger J. Flow evaluation software for four-dimensional flow MRI: a reliability and validation study. Radiol Med (Torino) 2023;128:1225–35. https://doi.org/10.1007/s11547-023-01697-4.

[44] Heiberg E, Green C, Toger J, Andersson AM, Carlsson M, Arheden H. FourFlow - open source code software for quantification and visualization of time-resolved three-directional phase contrast magnetic resonance velocity mapping. J Cardiovasc Magn Reson 2012;14:W14. https://doi.org/10.1186/1532-429X-14-S1-W14.
[45] Rizk J. 4D flow MRI applications in congenital heart disease. Eur Radiol 2021;31:1160–74. https://doi.org/10.1007/s00330-020-07210-z.
[46] Barker AJ, Markl M, Bürk J, Lorenz R, Bock J, Bauer S, et al. Bicuspid aortic valve is associated with altered wall shear stress in the ascending aorta. Circ Cardiovasc Imaging 2012;5:457–66. https://doi.org/10.1161/CIRCIMAGING.112.973370.
[47] Khiabani RH, Whitehead KK, Han D, Restrepo M, Tang E, Bethel J, et al. Exercise capacity in single-ventricle patients after Fontan correlates with haemodynamic energy loss in TCPC. Heart 2015;101:139–43. https://doi.org/10.1136/heartjnl-2014-306337.
[48] Rijnberg FM, Westenberg JJM, van Assen HC, Juffermans JF, Kroft LJM, van den Boogaard PJ, et al. 4D flow cardiovascular magnetic resonance derived energetics in the Fontan circulation correlate with exercise capacity and CMR-derived liver fibrosis/congestion. J Cardiovasc Magn Reson Off J Soc Cardiovasc Magn Reson 2022;24:21. https://doi.org/10.1186/s12968-022-00854-4.
[49] Fredriksson A, Trzebiatowska-Krzynska A, Dyverfeldt P, Engvall J, Ebbers T, Carlhäll C-J. Turbulent kinetic energy in the right ventricle: Potential MR marker for risk stratification of adults with repaired Tetralogy of Fallot. J Magn Reson Imaging JMRI 2018;47:1043–53. https://doi.org/10.1002/jmri.25830.
[50] Corral-Acero J, Margara F, Marciniak M, Rodero C, Loncaric F, Feng Y, et al. The “Digital Twin” to enable the vision of precision cardiology. Eur Heart J 2020;41:4556–64. https://doi.org/10.1093/eurheartj/ehaa159.
[51] Aslan S, Mangine NR, Laurence DW, Sabin PM, Wu W, Herz C, et al. Simulation of Transcatheter Therapies for Atrioventricular Valve Regurgitation in an Open-Source Finite Element Simulation Framework. Ann Biomed Eng 2026. https://doi.org/10.1007/s10439-025-03959-w.
[52] Zelonis CN, Maheshwari J, Wu W, Maas SA, Aslan S, Sunderland K, et al. Integrated Open-Source Framework for Simulation of Transcatheter Pulmonary Valves in Native Right Ventricular Outflow Tracts 2026. https://doi.org/10.48550/ARXIV.2507.06337.
[53] Karniadakis GE, Kevrekidis IG, Lu L, Perdikaris P, Wang S, Yang L. Physics-informed machine learning. Nat Rev Phys 2021;3:422–40. https://doi.org/10.1038/s42254-021-00314-5.
[54] Vinuesa R, Brunton SL. Enhancing computational fluid dynamics with machine learning. Nat Comput Sci 2022;2:358–66. https://doi.org/10.1038/s43588-022-00264-7.
[55] Pegolotti L, Pfaller MR, Rubio NL, Ding K, Brugarolas Brufau R, Darve E, et al. Learning reduced-order models for cardiovascular simulations with graph neural networks. Comput Biol Med 2024;168:107676. https://doi.org/10.1016/j.compbiomed.2023.107676.
[56] Raissi M, Yazdani A, Karniadakis GE. Hidden fluid mechanics: Learning velocity and pressure fields from flow visualizations. Science 2020;367:1026–30. https://doi.org/10.1126/science.aaw4741.
[57] Kissas G, Yang Y, Hwuang E, Witschey WR, Detre JA, Perdikaris P. Machine learning in cardiovascular flows modeling: Predicting arterial blood pressure from non-invasive 4D flow MRI data using physics-informed neural networks. Comput Methods Appl Mech Eng 2020;358:112623. https://doi.org/10.1016/j.cma.2019.112623.
[58] Fathi MF, Perez-Raya I, Baghaie A, Berg P, Janiga G, Arzani A, et al. Super-resolution and denoising of 4D-Flow MRI using physics-Informed deep neural nets. Comput Methods Programs Biomed 2020;197:105729. https://doi.org/10.1016/j.cmpb.2020.105729.
[59] Pfaff T, Fortunato M, Sanchez-Gonzalez A, Battaglia PW. Learning Mesh-Based Simulation with Graph Networks 2020. https://doi.org/10.48550/ARXIV.2010.03409.

[60] Bodnar C, Bruinsma WP, Lucic A, Stanley M, Allen A, Brandstetter J, et al. A foundation model for the Earth system. Nature 2025;641:1180–7. https://doi.org/10.1038/s41586-025-09005-y.
[61] Otsu N. A Threshold Selection Method from Gray-Level Histograms. IEEE Trans Syst Man Cybern 1979;9:62–6. https://doi.org/10.1109/TSMC.1979.4310076.
[62] Fischer C, Wetzl J, Schaeffter T, Giese D. Fully automated background phase correction using M-estimate SAmple consensus (MSAC)-Application to 2D and 4D flow. Magn Reson Med 2022;88:2709–17. https://doi.org/10.1002/mrm.29363.
[63] Beck A, Teboulle M. A Fast Iterative Shrinkage-Thresholding Algorithm for Linear Inverse Problems. SIAM J Imaging Sci 2009;2:183–202. https://doi.org/10.1137/080716542.
[64] Gatehouse PD, Rolf MP, Graves MJ, Hofman MB, Totman J, Werner B, et al. Flow measurement by cardiovascular magnetic resonance: a multi-centre multi-vendor study of background phase offset errors that can compromise the accuracy of derived regurgitant or shunt flow measurements. J Cardiovasc Magn Reson Off J Soc Cardiovasc Magn Reson 2010;12:5. https://doi.org/10.1186/1532-429X-12-5.
[65] Schofield MA, Zhu Y. Fast phase unwrapping algorithm for interferometric applications. Opt Lett 2003;28:1194–6. https://doi.org/10.1364/ol.28.001194.
[66] Loecher M, Schrauben E, Johnson KM, Wieben O. Phase unwrapping in 4D MR flow with a 4D single-step laplacian algorithm. J Magn Reson Imaging JMRI 2016;43:833–42. https://doi.org/10.1002/jmri.25045.
[67] Bock J, Frydrychowicz A, Stalder AF, Bley TA, Burkhardt H, Hennig J, et al. 4D phase contrast MRI at 3 T: effect of standard and blood-pool contrast agents on SNR, PC-MRA, and blood flow visualization. Magn Reson Med 2010;63:330–8. https://doi.org/10.1002/mrm.22199.
[68] Song SM, Napel S, Glover GH, Pelc NJ. Noise reduction in three-dimensional phase-contrast MR velocity measurements. J Magn Reson Imaging JMRI 1993;3:587–96. https://doi.org/10.1002/jmri.1880030407.
[69] Loecher M, Kecskemeti S, Turski P, Wieben O. Comparison of divergence-free algorithms for 3D MRI with three-directional velocity encoding. J Cardiovasc Magn Reson 2012;14:W64. https://doi.org/10.1186/1532-429X-14-S1-W64.
[70] Mura J, Pino AM, Sotelo J, Valverde I, Tejos C, Andia ME, et al. Enhancing the Velocity Data From 4D Flow MR Images by Reducing its Divergence. IEEE Trans Med Imaging 2016;35:2353–64. https://doi.org/10.1109/TMI.2016.2570010.
[71] Ong F, Uecker M, Tariq U, Hsiao A, Alley MT, Vasanawala SS, et al. Robust 4D flow denoising using divergence-free wavelet transform. Magn Reson Med 2015;73:828–42. https://doi.org/10.1002/mrm.25176.
[72] Bostan E, Unser M, Ward JP. Divergence-Free Wavelet Frames. IEEE Signal Process Lett 2015;22:1142–6. https://doi.org/10.1109/LSP.2015.2388794.
[73] Starck J-L, Fadili J, Murtagh F. The undecimated wavelet decomposition and its reconstruction. IEEE Trans Image Process Publ IEEE Signal Process Soc 2007;16:297–309. https://doi.org/10.1109/tip.2006.887733.
[74] Knutsson H, Westin C-F. Normalized and differential convolution. Proc. IEEE Conf. Comput. Vis. Pattern Recognit., New York, NY, USA: IEEE Comput. Soc. Press; 1993, p. 515–23. https://doi.org/10.1109/CVPR.1993.341081.
[75] Antiga L, Piccinelli M, Botti L, Ene-Iordache B, Remuzzi A, Steinman DA. An image-based modeling framework for patient-specific computational hemodynamics. Med Biol Eng Comput 2008;46:1097–112. https://doi.org/10.1007/s11517-008-0420-1.
[76] Wigström L, Ebbers T, Fyrenius A, Karlsson M, Engvall J, Wranne B, et al. Particle trace visualization of intracardiac flow using time-resolved 3D phase contrast MRI. Magn Reson Med 1999;41:793–9. https://doi.org/10.1002/(sici)1522-2594(199904)41:4%3C793::aid-mrm19%3E3.0.co;2-2.

[77] Lorenz R, Bock J, Barker AJ, von Knobelsdorff-Brenkenhoff F, Wallis W, Korvink JG, et al. 4D flow magnetic resonance imaging in bicuspid aortic valve disease demonstrates altered distribution of aortic blood flow helicity. Magn Reson Med 2014;71:1542–53. https://doi.org/10.1002/mrm.24802.
[78] Elbaz MSM, Calkoen EE, Westenberg JJM, Lelieveldt BPF, Roest AAW, van der Geest RJ. Vortex flow during early and late left ventricular filling in normal subjects: quantitative characterization using retrospectively-gated 4D flow cardiovascular magnetic resonance and three-dimensional vortex core analysis. J Cardiovasc Magn Reson Off J Soc Cardiovasc Magn Reson 2014;16:78. https://doi.org/10.1186/s12968-014-0078-9.
[79] Carlsson M, Heiberg E, Toger J, Arheden H. Quantification of left and right ventricular kinetic energy using four-dimensional intracardiac magnetic resonance imaging flow measurements. Am J Physiol Heart Circ Physiol 2012;302:H893-900. https://doi.org/10.1152/ajpheart.00942.2011.
[80] Barker AJ, van Ooij P, Bandi K, Garcia J, Albaghdadi M, McCarthy P, et al. Viscous energy loss in the presence of abnormal aortic flow. Magn Reson Med 2014;72:620–8. https://doi.org/10.1002/mrm.24962.
[81] Lorensen WE, Cline HE. Marching cubes: A high resolution 3D surface construction algorithm. ACM SIGGRAPH Comput Graph 1987;21:163–9. https://doi.org/10.1145/37402.37422.
[82] Ku DN, Giddens DP, Zarins CK, Glagov S. Pulsatile flow and atherosclerosis in the human carotid bifurcation. Positive correlation between plaque location and low oscillating shear stress. Arteriosclerosis 1985;5:293–302. https://doi.org/10.1161/01.atv.5.3.293.
[83] Stalder AF, Russe MF, Frydrychowicz A, Bock J, Hennig J, Markl M. Quantitative 2D and 3D phase contrast MRI: optimized analysis of blood flow and vessel wall parameters. Magn Reson Med 2008;60:1218–31. https://doi.org/10.1002/mrm.21778.
[84] Potters WV, van Ooij P, Marquering H, vanBavel E, Nederveen AJ. Volumetric arterial wall shear stress calculation based on cine phase contrast MRI. J Magn Reson Imaging JMRI 2015;41:505–16. https://doi.org/10.1002/jmri.24560.
[85] Petersson S, Dyverfeldt P, Ebbers T. Assessment of the accuracy of MRI wall shear stress estimation using numerical simulations. J Magn Reson Imaging JMRI 2012;36:128–38. https://doi.org/10.1002/jmri.23610.
[86] Ebbers T, Wigström L, Bolger AF, Engvall J, Karlsson M. Estimation of relative cardiovascular pressures using time-resolved three-dimensional phase contrast MRI. Magn Reson Med 2001;45:872–9. https://doi.org/10.1002/mrm.1116.
[87] Tyszka JM, Laidlaw DH, Asa JW, Silverman JM. Three-dimensional, time-resolved (4D) relative pressure mapping using magnetic resonance imaging. J Magn Reson Imaging JMRI 2000;12:321–9. https://doi.org/10.1002/1522-2586(200008)12:2%3C321::aid-jmri15%3E3.0.co;2-2.
[88] Ebbers T, Farnebäck G. Improving computation of cardiovascular relative pressure fields from velocity MRI. J Magn Reson Imaging JMRI 2009;30:54–61. https://doi.org/10.1002/jmri.21775.
[89] Krittian SBS, Lamata P, Michler C, Nordsletten DA, Bock J, Bradley CP, et al. A finite-element approach to the direct computation of relative cardiovascular pressure from time-resolved MR velocity data. Med Image Anal 2012;16:1029–37. https://doi.org/10.1016/j.media.2012.04.003.
[90] Bock J, Frydrychowicz A, Lorenz R, Hirtler D, Barker AJ, Johnson KM, et al. In vivo noninvasive 4D pressure difference mapping in the human aorta: phantom comparison and application in healthy volunteers and patients. Magn Reson Med 2011;66:1079–88. https://doi.org/10.1002/mrm.22907.

[91] Bernstein MA, Zhou XJ, Polzin JA, King KF, Ganin A, Pelc NJ, et al. Concomitant gradient terms in phase contrast MR: analysis and correction. Magn Reson Med 1998;39:300–8. https://doi.org/10.1002/mrm.1910390218.

## Graphical Abstract

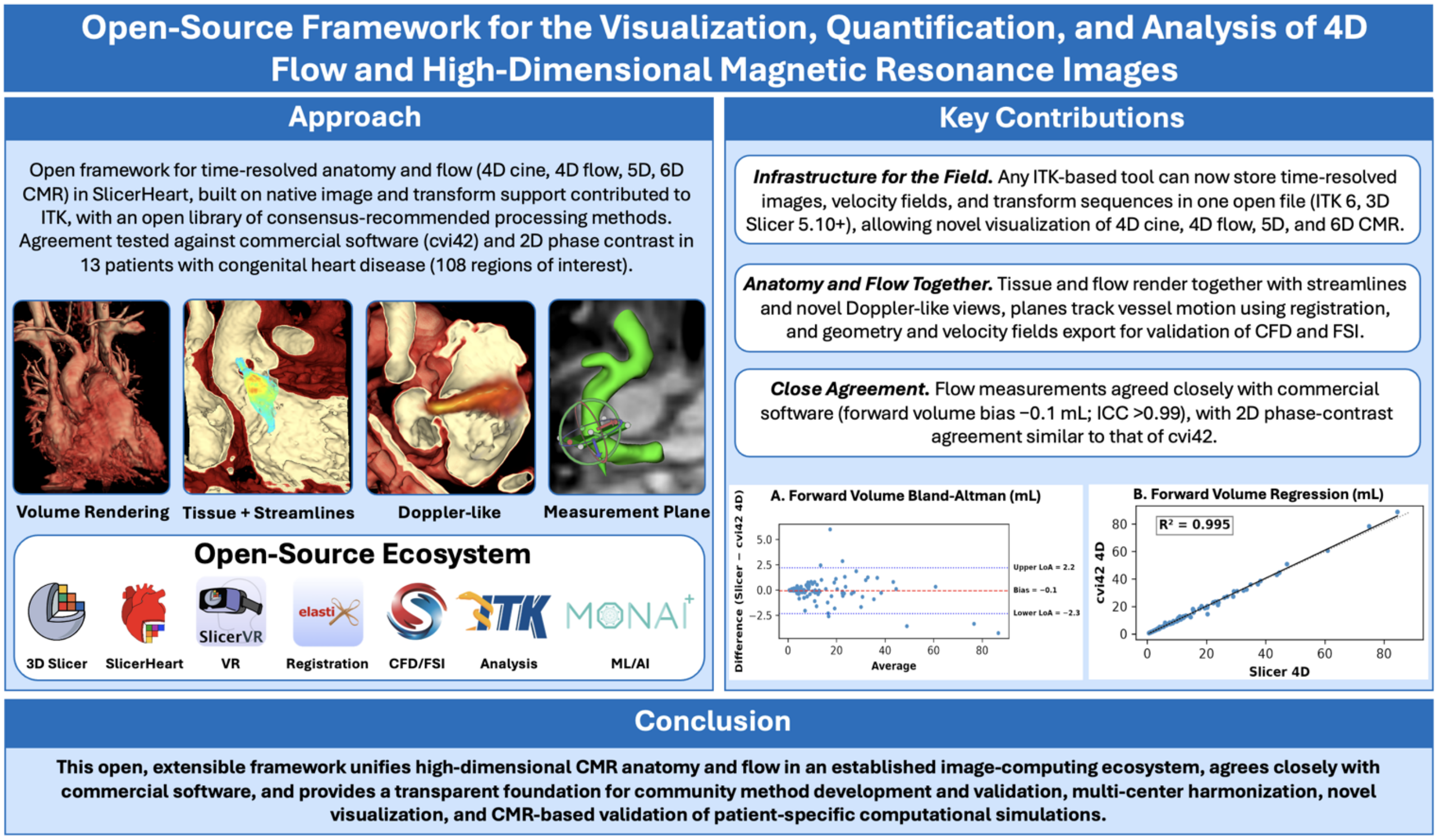

## Tables

### Table 1. Patient demographics and clinical characteristics.

| Patient | Sex | Age (years) | Diagnosis | Prior Intervention | ROIs |
|---|---|---|---|---|---|
| 1 | Male | 17.1 | Sinus venosus ASD, after Warden repair | Yes | 8 |
| 2 | Male | 3.9 | Ebstein anomaly with significant tricuspid regurgitation | Yes (Cath) | 7 |
| 3 | Female | 3.4 | ccTGA {I,D,D} with VSD, single-ventricle palliation pathway | Yes | 9 |
| 4 | Male | 5.6 | Restrictive conoventricular VSD | No | 8 |
| 5 | Female | 5.6 | Taussig-Bing DORV, after arterial switch and VSD closure | Yes | 7 |
| 6 | Female | 10.8 | Conoventricular VSD, after repair with residual tricuspid regurgitation | Yes | 7 |
| 7 | Male | 0.4 | VSD with branch pulmonary artery hypoplasia | No | 5 |
| 8 | Female | 0.4 | Supracardiac TAPVC, after sutureless repair | Yes | 9 |
| 9 | Male | 4.5 | PA/IVS, after repair with subsequent interventions | Yes | 10 |
| 10 | Female | 0.2 | Unobstructed supracardiac TAPVC, after repair | Yes | 12 |
| 11 | Female | 0.6 | Heterotaxy/polysplenia, single ventricle, after Kawashima | Yes | 9 |
| 12 | Male | 0.04 | DORV | No | 9 |
| 13 | Female | 8.1 | Residual ASD after device closure | Yes (Cath) | 8 |

Thirteen patients with congenital heart disease (6 male, 7 female) were included, ranging in age from 0.04 to 17.1 years (median 3.9 years, IQR 0.4 to 5.6). Age is expressed in decimal years. Ten patients had undergone prior surgical or catheter-based intervention, two by catheter only (Cath). ROIs indicate the number of regions of interest (vessel segments) per patient with paired framework and cvi42 4D flow measurements, totaling 108 ROIs. 4D = four-dimensional; ASD = atrial septal defect; Cath = catheter-based intervention; ccTGA = congenitally corrected transposition of the great arteries; DORV = double outlet right ventricle; IQR = interquartile range; PA/IVS = pulmonary atresia with intact ventricular septum; ROI = region of interest; TAPVC = total anomalous pulmonary venous connection; VSD = ventricular septal defect.

**Table 2. Agreement between the framework (3D Slicer) and cvi42 4D flow for hemodynamic metrics, by Bland-Altman analysis across 108 paired ROIs in 13 patients.**

| Metric | n | Bias (95% CI) | 95% LoA* | MAPE | ICC (95% CI) |
|---|---|---|---|---|---|
| Forward volume (mL)† | 108 | −0.1 (−0.3 to 0.1) | 2.2 to −2.3 | 4.8% | 0.997 (0.990–0.999) |
| Effective volume per minute (L/min)† | 108 | −0.015 (−0.038 to 0.004) | 0.21 to −0.24 | 6.4% | 0.994 (0.988–0.997) |
| Mean velocity (cm/s) | 108 | −2.0 (−4.8 to −0.6) | 9.8 to −13.9 | 9.2% | 0.795 (0.560–0.973) |
| Maximum flow (mL/s) | 108 | 1.2 (−0.5 to 2.8) | 16.8 to −14.4 | 9.0% | 0.993 (0.981–0.998) |
| Cross-sectional area ($mm^2$) | 108 | 5.1 (1.5 to 8.5) | 46.5 to −36.3 | 11.1% | 0.986 (0.972–0.991) |

Bias is the mean difference (framework minus cvi42). *Limits of agreement adjusted for multiple ROIs per patient [42]; their confidence intervals are reported in Supplementary Table S1. †Primary endpoints. MAPE was calculated relative to the cvi42 value. ICCs (two-way mixed effects, absolute agreement, single measurement) quantify absolute agreement between methods. Confidence intervals for bias and ICC were estimated by patient-level cluster bootstrap (20,000 resamples). 4D = four-dimensional; CI = confidence interval; ICC = intraclass correlation coefficient; LoA = limits of agreement; MAPE = mean absolute percentage error; ROI = region of interest.

## Figures

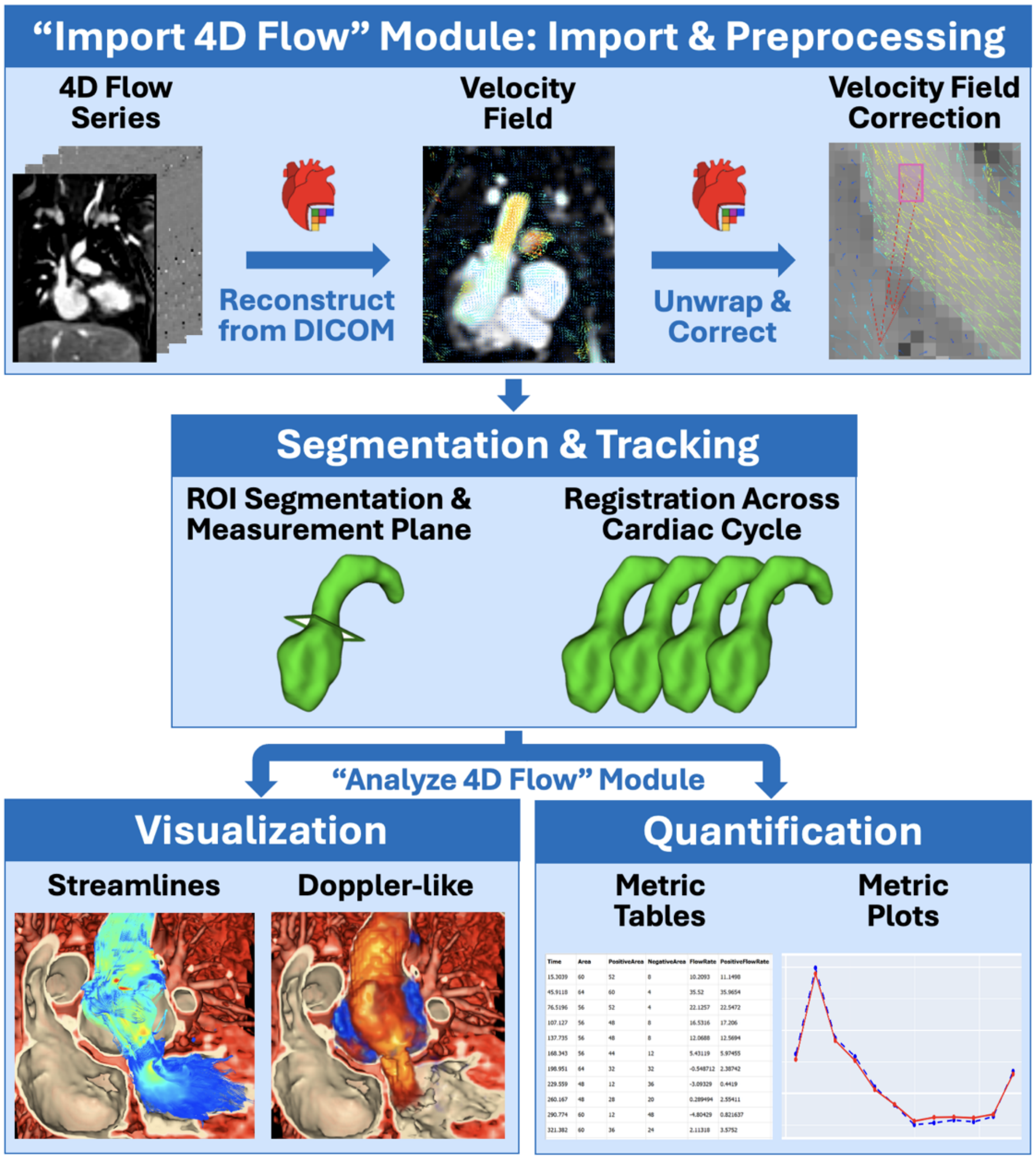

**Figure 1.** Workflow of the 4D flow framework in SlicerHeart. The "Import 4D Flow" module reconstructs the 4D flow series from DICOM data into a velocity field and corrects phase wraps (aliasing) and background phase offset; the magenta box marks an aliased velocity vector (red) identified for correction. The vessel ROI is segmented with a measurement plane and propagated across the cardiac cycle by deformable registration. The "Analyze 4D Flow" module displays flow as streamlines and Doppler-like renderings within the anatomy and quantifies flow metrics in tables and plots (examples for illustration). The workflow is demonstrated in Movies 1 and 2. 4D = four-dimensional; DICOM = Digital Imaging and Communications in Medicine; ROI = region of interest.

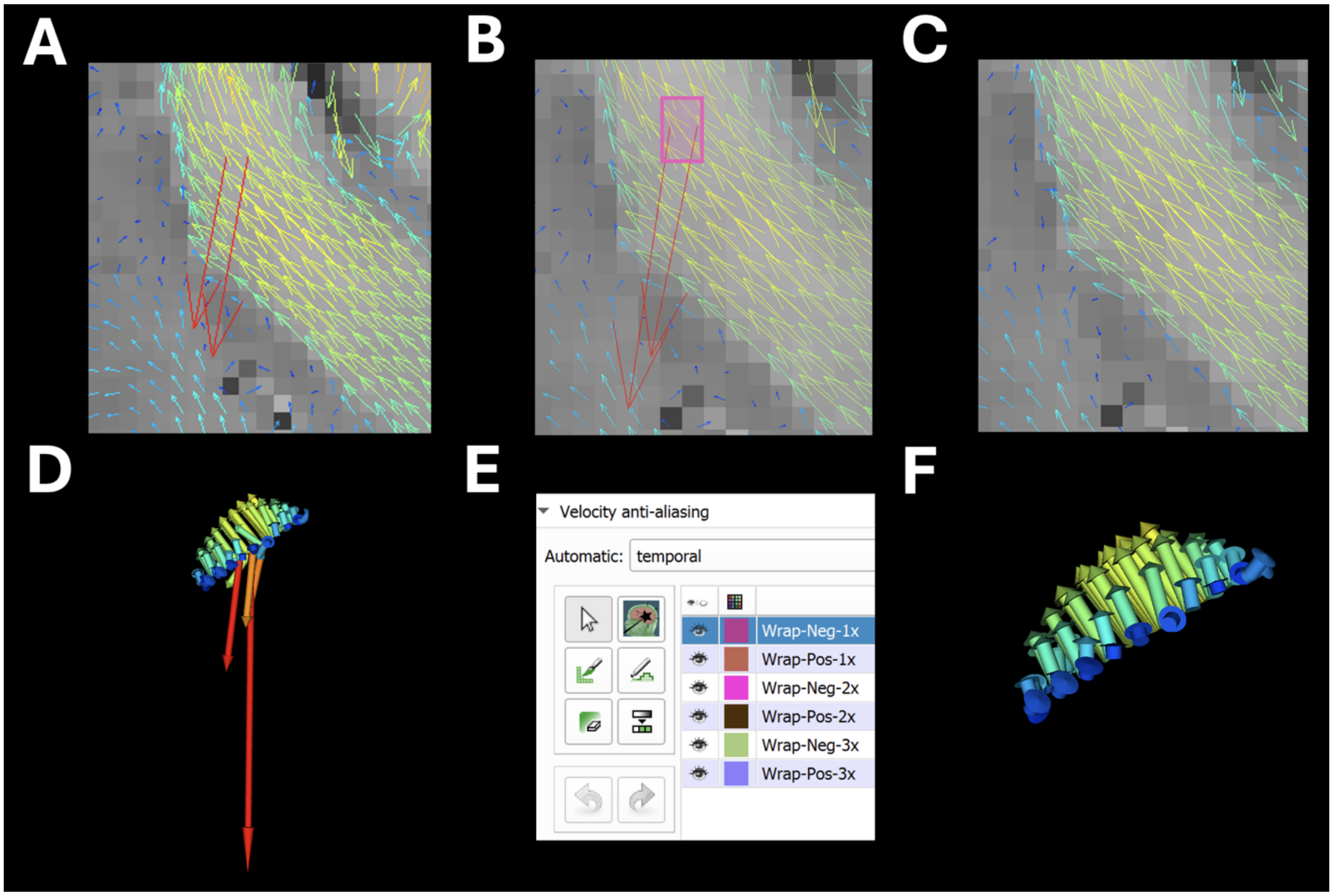

**Figure 2.** Correction of velocity aliasing by phase unwrapping. (A–C) Two-dimensional sagittal reformatted views of 4D flow velocity vectors in the aorta: (A) aliased vectors (red) pointing opposite to the surrounding flow; (B) the aliased voxel identified for correction (magenta box); (C) the velocity field after unwrapping. (D) Three-dimensional glyph view of the aliased vectors, colored by velocity magnitude. (E) Velocity anti-aliasing interface: the automatic method (temporal) generates wrap-count labels for up to three successive wraps in either direction (Wrap-Neg-1x to Wrap-Pos-3x), which can be reviewed and edited before the corrected velocities are regenerated. (F) Three-dimensional glyph view after unwrapping. 4D = four-dimensional.

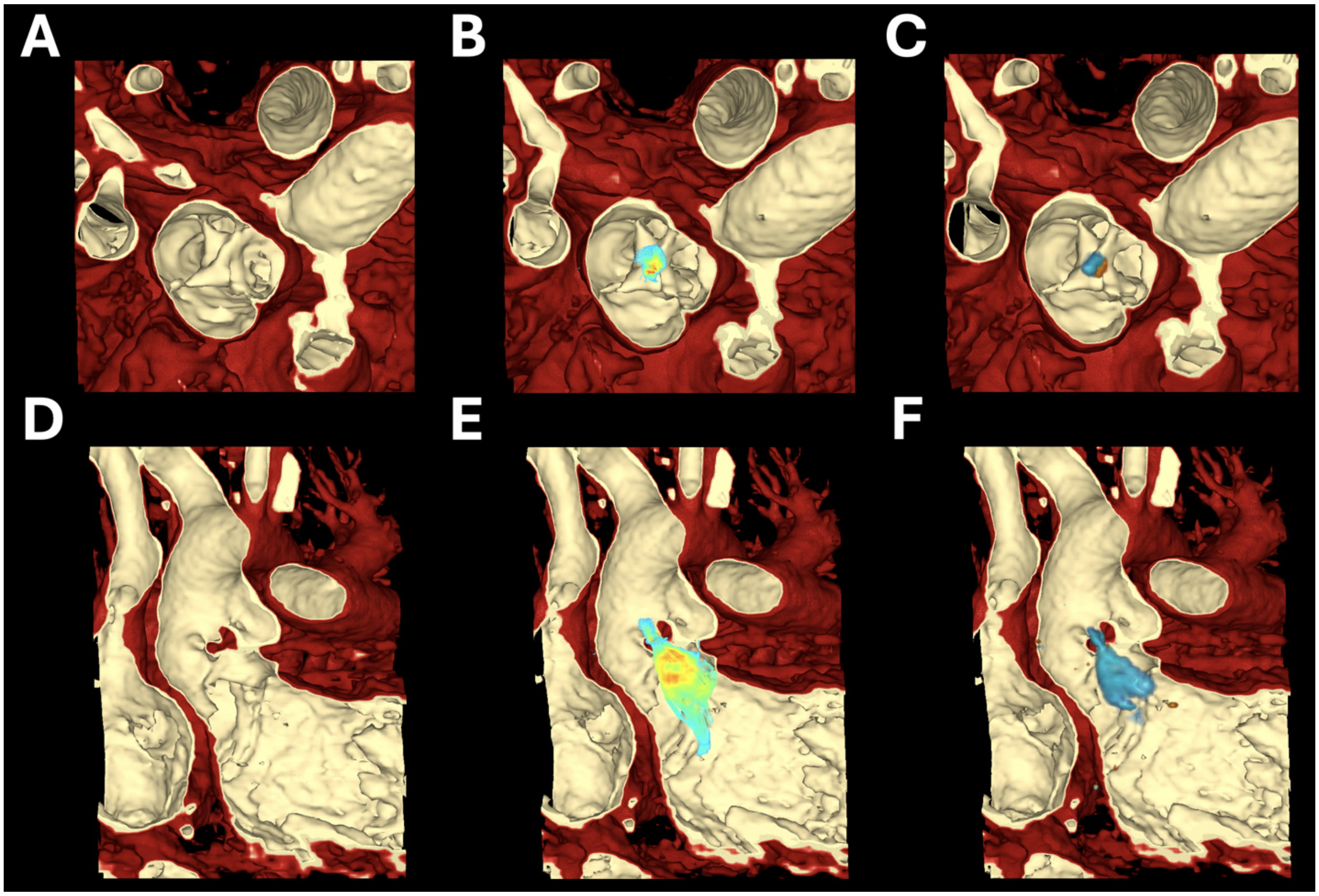

**Figure 3.** Combined rendering of tissue and flow in diastole from 4D flow CMR in a 19-year-old male with truncus arteriosus after repair with VSD closure, RV-PA conduit, and pulmonary valve replacement (illustrative case). Top row (A–C), surgical view; bottom row (D–F), coronal view of the quadricuspid truncal valve. From left to right: tissue alone (A, D), tissue with regurgitant streamlines (B, E), and tissue with a Doppler-like rendering of velocity relative to the annular plane (C, F). Additional imaging of this patient is shown in Movie 3. 4D = four-dimensional; CMR = cardiovascular magnetic resonance; RV-PA = right ventricle-to-pulmonary artery; VSD = ventricular septal defect.

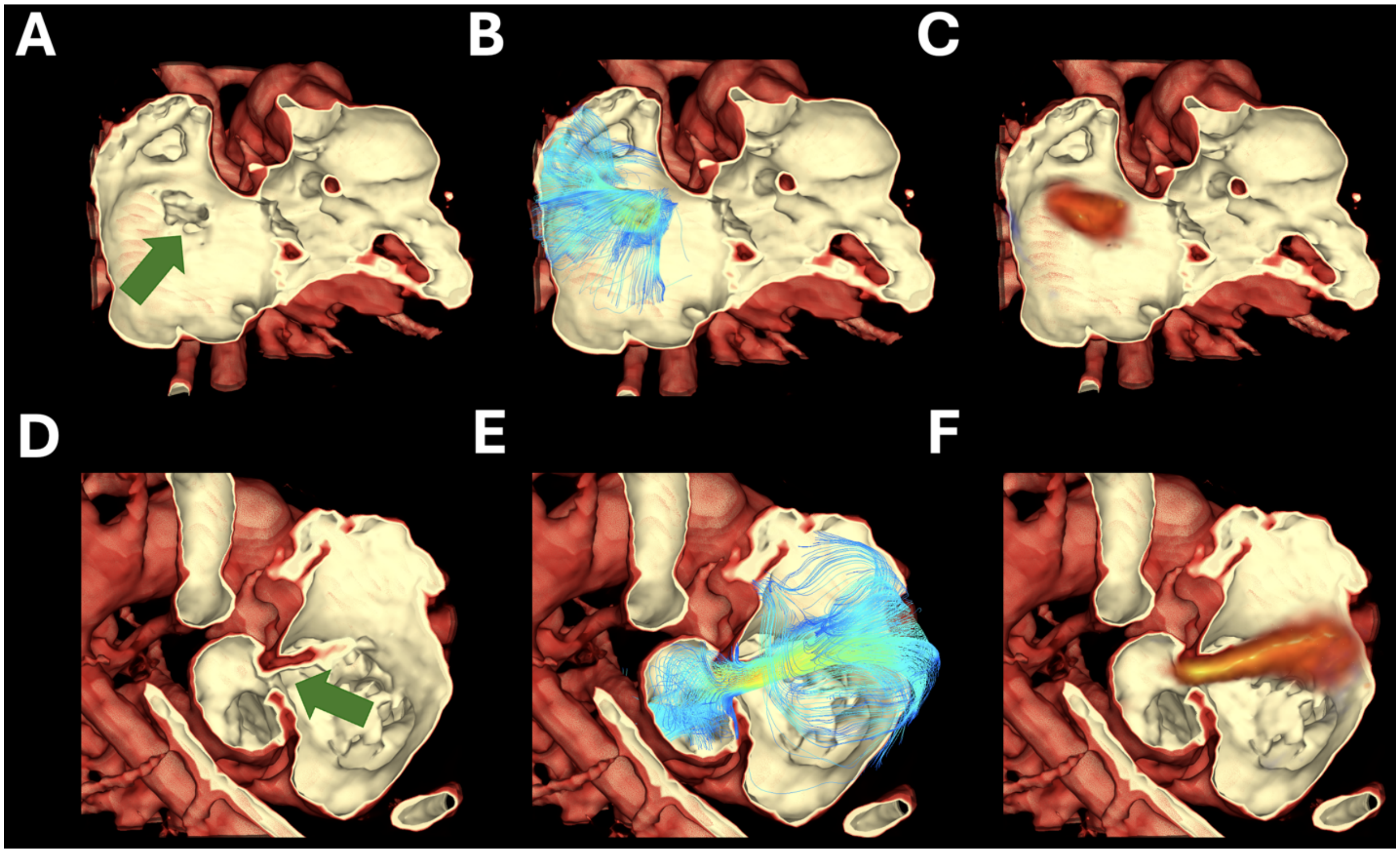

**Figure 4.** Combined rendering of tissue and flow in systole from 4D flow CMR in a 3-year-old female with hypoplastic left heart syndrome after bidirectional Glenn with left pulmonary arterioplasty, Sano shunt ligation, and stent angioplasty of a restrictive atrial septum (illustrative case). Top row (A–C), coronal view; bottom row (D–F), short-axis view at the atrial level. From left to right: tissue alone, with the atrial septal communication marked (green arrow) (A, D); tissue with atrial shunt streamlines (B, E); and tissue with a Doppler-like rendering of velocity relative to the septal plane (C, F). Additional imaging of this patient is shown in Movie 5. 4D = four-dimensional; CMR = cardiovascular magnetic resonance.

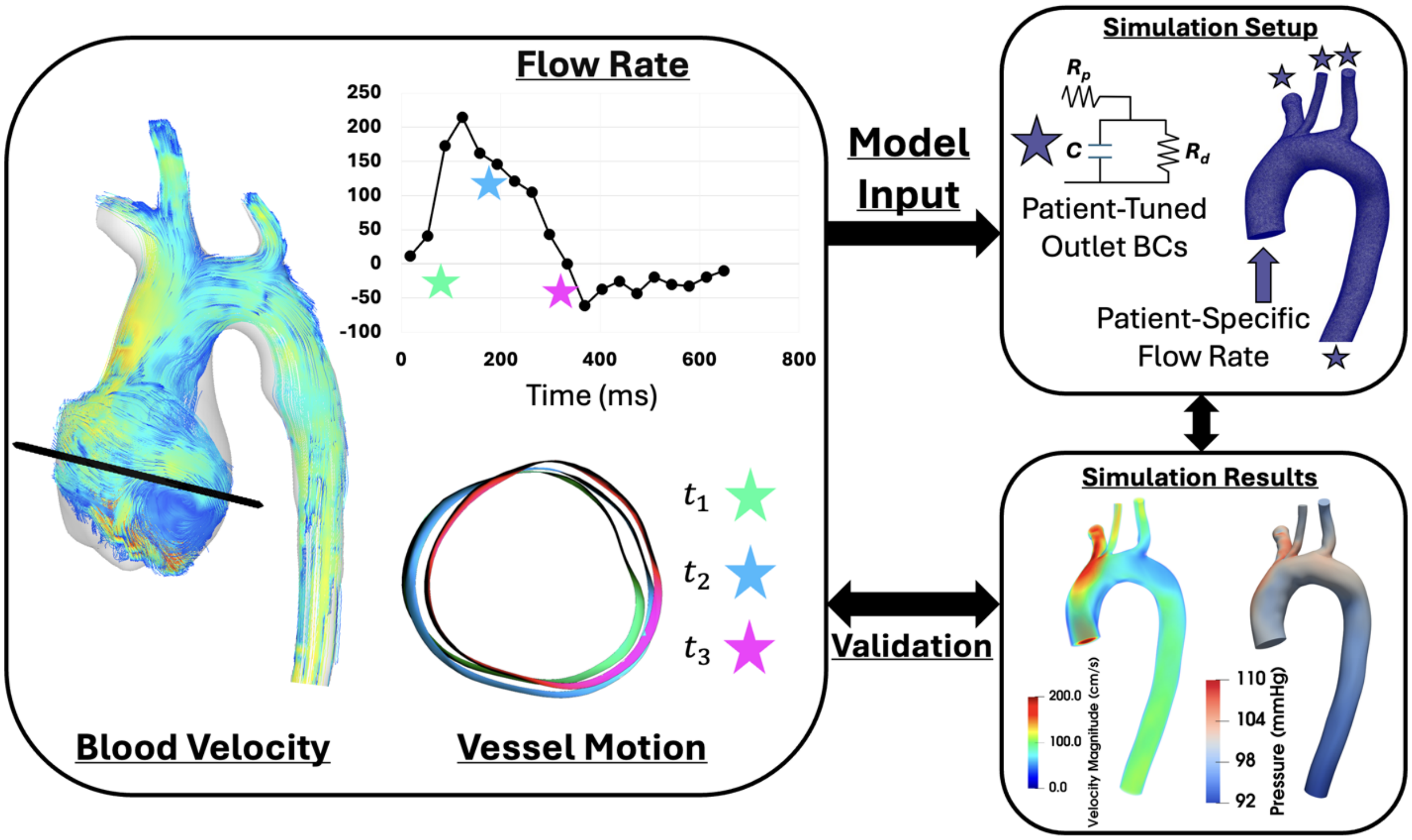

**Figure 5.** 4D CMR flow and motion data to inform validation and accuracy of patient-specific physics-based simulations (computational fluid dynamics and fluid-structure interaction). 4D = four-dimensional; BCs = boundary conditions; C = compliance; CMR = cardiovascular magnetic resonance; $R_d$ = distal resistance; $R_p$ = proximal resistance.

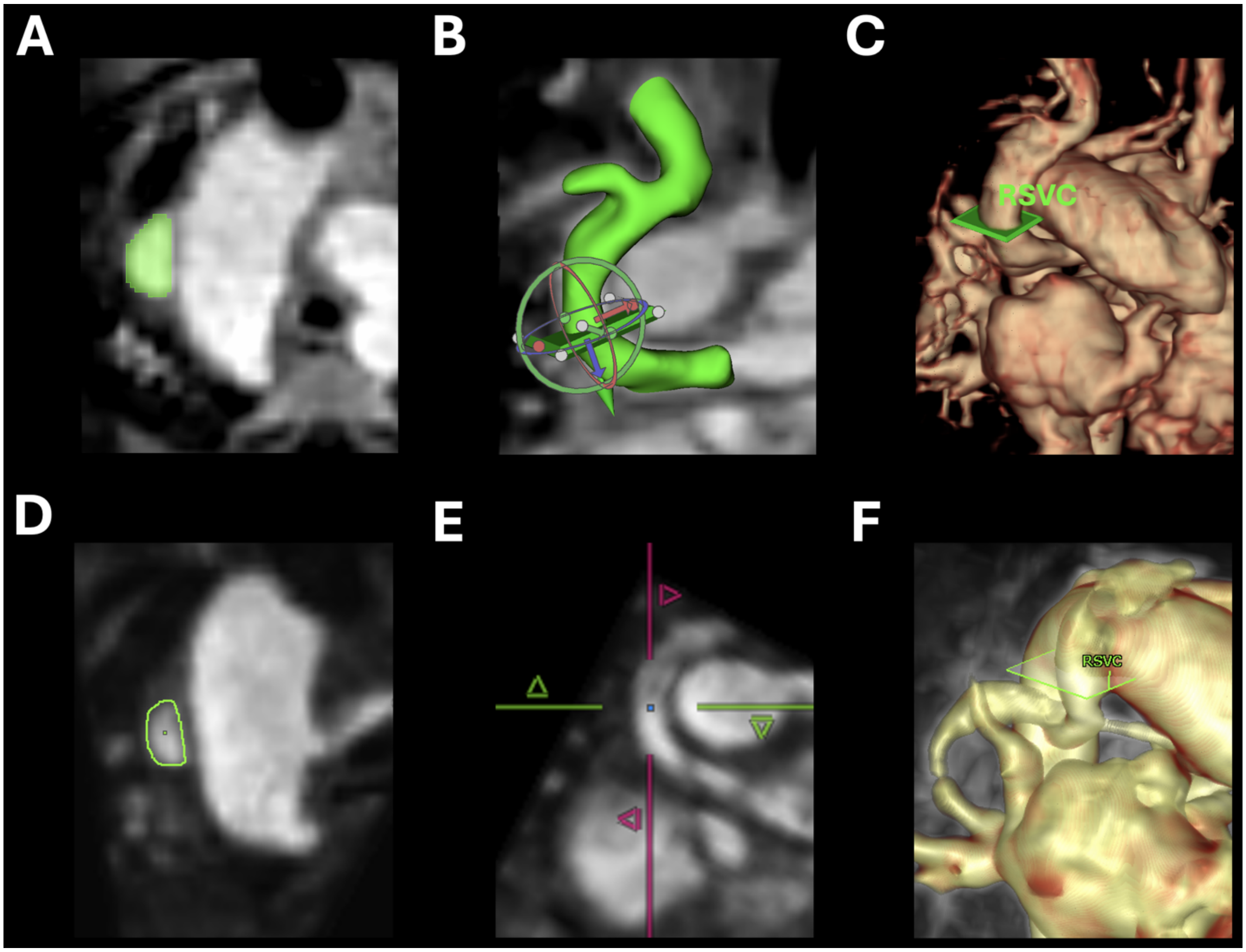

**Figure 6.** Segmentation and measurement plane placement in the framework (A–C) and cvi42 (D–F), from 4D flow CMR in a 7-month-old female with heterotaxy, single-ventricle physiology, and bilateral superior venae cavae. From left to right: RSVC segmentation (green) on a cross-sectional image (A, D); measurement plane positioned perpendicular to the vessel, with interaction handles (B) or reference line (magenta) (E); and volume rendering with the labeled RSVC plane (C, F). 4D = four-dimensional; CMR = cardiovascular magnetic resonance; RSVC = right superior vena cava.

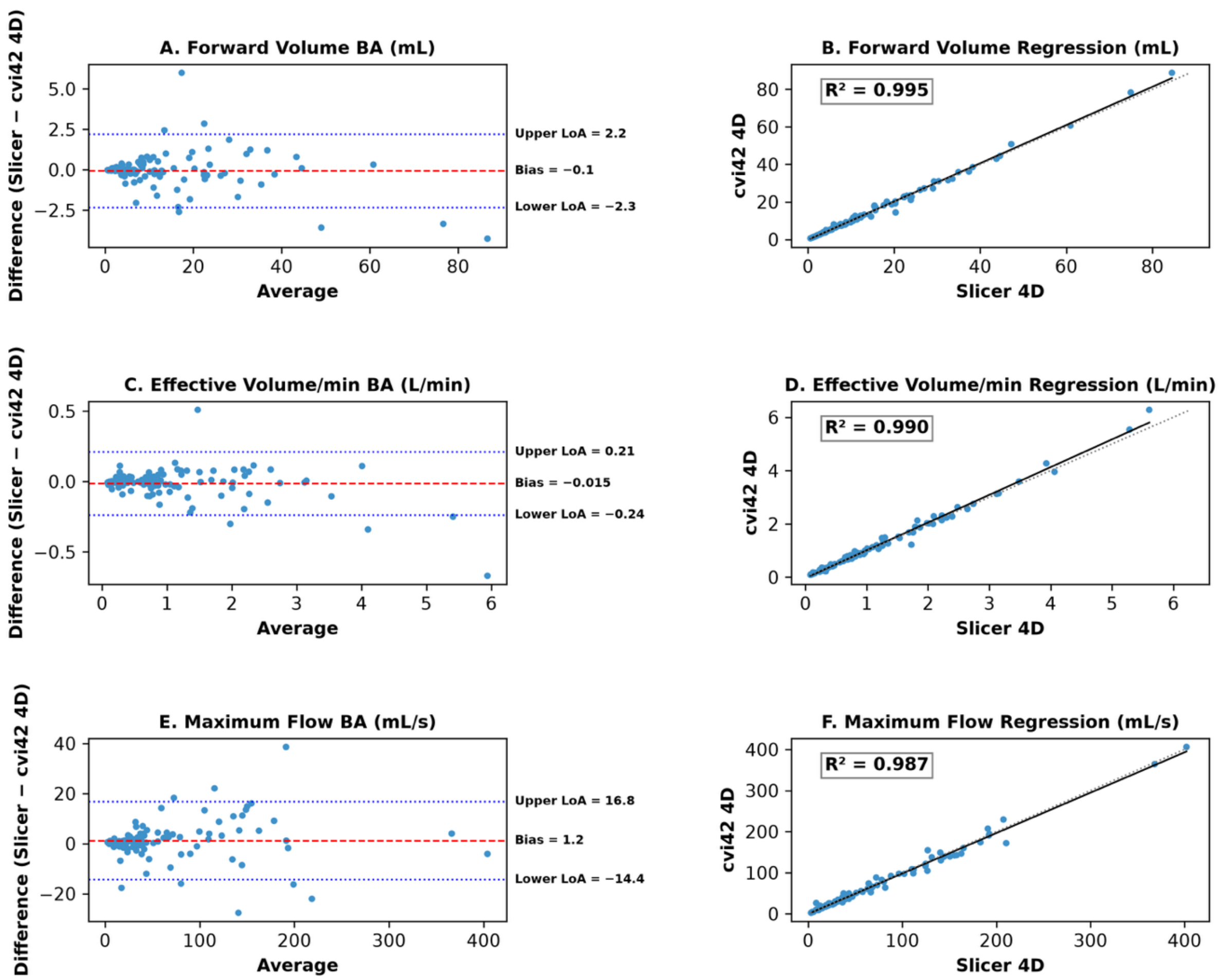

**Figure 7.** Agreement between the framework (3D Slicer) and cvi42 4D flow measurements across 108 paired ROIs in 13 patients. Bland-Altman plots (left) and regression of cvi42 against framework values (right) for forward volume (A, B), effective volume per minute (C, D), and maximum flow (E, F). In Bland-Altman plots, the difference (framework minus cvi42) is plotted against the mean of the two methods; red dashed lines indicate the bias and blue dotted lines the 95% LoA, adjusted for multiple ROIs per patient. In regression plots, black solid lines show the least-squares fit and gray dotted lines the line of identity. All ROIs are shown. 4D = four-dimensional; BA = Bland-Altman; LoA = limits of agreement; $R^2$ = coefficient of determination; ROI = region of interest.

## Movie Legends

**Movie 1.** Retrieval and import of 4D flow CMR data into 3D Slicer. 4D flow series are retrieved from the PACS, organized, and imported with the “Import 4D Flow” module for visualization and analysis. 4D = four-dimensional; CMR = cardiovascular magnetic resonance; PACS = picture archiving and communication system.

**Movie 2.** 4D flow processing and quantification in SlicerHeart, including phase unwrapping, anatomic segmentation, measurement plane placement, velocity glyph and streamline visualization, and automated quantification of hemodynamic metrics. 4D = four-dimensional.

**Movie 3.** Free-running cardiac- and respiratory-motion-resolved 5D MRA and 4D flow CMR in a 19-year-old male with truncus arteriosus after repair with VSD closure, RV-PA conduit, and pulmonary valve replacement (illustrative case; same patient as Figure 3). Shown in sequence: (1) volume rendering of the truncal valve from 5D MRA (single respiratory phase) in coronal and axial views; (2) 4D flow CMR with tissue rendering; and (3) tissue rendering with streamlines and with Doppler-like rendering, showing regurgitation through the truncal valve coaptation defect. 4D = four-dimensional; 5D = five-dimensional; CMR = cardiovascular magnetic resonance; MRA = magnetic resonance angiography; RV-PA = right ventricle-to-pulmonary artery; VSD = ventricular septal defect.

**Movie 4.** Free-running cardiac- and respiratory-motion-resolved 5D MRA and 4D flow CMR in a 5-year-old male with unbalanced atrioventricular canal defect (unbalanced to the left) after atrioventricular valve repair (illustrative case). Shown in sequence: (1) volume rendering from 5D MRA (single respiratory phase) in coronal and axial views; (2) 4D flow CMR with tissue rendering; and (3) tissue rendering with streamlines and with Doppler-like rendering, showing systolic atrioventricular valve regurgitation. 4D = four-dimensional; 5D = five-dimensional; CMR = cardiovascular magnetic resonance; MRA = magnetic resonance angiography.

**Movie 5.** 4D flow CMR in a 3-year-old female with hypoplastic left heart syndrome after bidirectional Glenn with left pulmonary arterioplasty, Sano shunt ligation, and stent angioplasty of a restrictive atrial septum (illustrative case; same patient as Figure 4). Volume rendering from 4D flow CMR with the stented atrial septal communication marked (red arrow), followed by tissue rendering with streamlines and with Doppler-like rendering, showing flow across the atrial septal communication. 4D = four-dimensional; CMR = cardiovascular magnetic resonance.

## Supplementary Material

### Supplementary Table S1. Agreement between the framework (3D Slicer) and cvi42 4D flow for all hemodynamic metrics, by Bland-Altman analysis across 108 paired ROIs in 13 patients.

| Metric | n | Bias (95% CI) | 95% LoA* | Upper LoA (95% CI) | Lower LoA (95% CI) | MAPE | ICC (95% CI) |
|---|---|---|---|---|---|---|---|
| Forward volume (mL)† | 108 | −0.1 (−0.3 to 0.1) | 2.2 to −2.3 | 1.4 to 3.0 | −3.1 to −1.5 | 4.8% | 0.997 (0.990–0.999) |
| Effective volume per minute (L/min)† | 108 | −0.015 (−0.038 to 0.004) | 0.21 to −0.24 | 0.12 to 0.28 | −0.34 to −0.13 | 6.4% | 0.994 (0.988–0.997) |
| Mean velocity (cm/s)† | 108 | −2.0 (−4.8 to −0.6) | 9.8 to −13.9 | 2.6 to 15.1 | −24.3 to −4.2 | 9.2% | 0.795 (0.560–0.973) |
| Maximum flow (mL/s)† | 108 | 1.2 (−0.5 to 2.8) | 16.8 to −14.4 | 12.0 to 21.5 | −18.1 to −10.3 | 9.0% | 0.993 (0.981–0.998) |
| Cross-sectional area ($mm^2$)† | 108 | 5.1 (1.5 to 8.5) | 46.5 to −36.3 | 33.0 to 58.4 | −47.8 to −24.5 | 11.1% | 0.986 (0.972–0.991) |
| Backward volume (mL) | 108 | −0.12 (−0.24 to −0.002) | 1.17 to −1.41 | 0.56 to 1.80 | −2.19 to −0.67 | 2.5% | 0.950 (0.709–0.989) |
| Total volume (mL) | 108 | −0.2 (−0.5 to 0.0) | 2.6 to −2.9 | 1.2 to 3.7 | −4.4 to −1.4 | 6.1% | 0.995 (0.989–0.997) |
| Net positive volume (mL) | 108 | 1.1 (0.6 to 1.6) | 5.1 to −2.9 | 3.8 to 6.1 | −3.2 to −2.4 | 10.1% | 0.991 (0.975–0.995) |
| Net negative volume (mL) | 108 | −1.3 (−2.0 to −0.7) | 3.2 to −5.8 | 1.9 to 4.3 | −8.1 to −3.5 | 9.8% | 0.811 (0.551–0.924) |
| Regurgitation fraction (%) | 108 | 0.8 (−0.9 to 1.9) | 11.4 to −9.7 | 6.4 to 15.1 | −15.9 to −4.1 | 2.5‡ | 0.915 (0.712–0.977) |
| Volume per minute (L/min) | 108 | −0.005 (−0.028 to 0.014) | 0.19 to −0.20 | 0.13 to 0.25 | −0.26 to −0.13 | 5.0% | 0.996 (0.989–0.998) |
| Peak velocity (cm/s) | 108 | −10.0 (−15.7 to −5.9) | 30.7 to −50.8 | 20.5 to 38.7 | −68.5 to −33.6 | 12.7% | 0.836 (0.769–0.908) |
| Maximum pressure gradient (mmHg) | 108 | −0.8 (−1.4 to −0.4) | 3.4 to −5.0 | 1.5 to 5.1 | −7.7 to −2.4 | 22.4% | 0.765 (0.654–0.904) |

Bias is the mean difference (framework minus cvi42). *Limits of agreement adjusted for multiple ROIs per patient [42]. †Metrics also reported in Table 2. MAPE was calculated relative to the cvi42 value; for backward, total, and net volumes, relative to the cvi42 forward volume. ‡For regurgitation fraction, the mean absolute difference in percentage points is shown instead of MAPE, because the cvi42 value can be zero. When cvi42 reported no backward flow, its regurgitation fraction was taken as 0%, and the framework value was compared with 0%. Confidence intervals were estimated by patient-level cluster bootstrap (20,000 resamples). 4D = four-dimensional; CI = confidence interval; ICC = intraclass correlation coefficient; LoA = limits of agreement; MAPE = mean absolute percentage error; ROI = region of interest.

**Supplementary Table S2. Agreement of the framework (labeled “Slicer”) and cvi42 4D flow with 2D phase contrast, by Bland-Altman analysis across the 100 paired ROIs with a 2D phase-contrast forward volume in 13 patients.**

| Metric | Comparison | n | Bias (95% CI) | 95% LoA* | Upper LoA (95% CI) | Lower LoA (95% CI) | MAPE | ICC (95% CI) |
|---|---|---|---|---|---|---|---|---|
| Forward volume (mL) | Slicer vs. 2D | 100 | −2.2 (−3.2 to −1.4) | 2.7 to −7.2 | 1.6 to 3.6 | −9.2 to −5.1 | 21.8% | 0.978 (0.934–0.991) |
| | cvi42 4D vs. 2D | 100 | −2.1 (−3.0 to −1.3) | 2.6 to −6.8 | 1.5 to 3.5 | −8.8 to −4.8 | 21.3% | 0.980 (0.940–0.992) |
| Backward volume (mL) | Slicer vs. 2D | 99 | 0.28 (−0.03 to 0.64) | 3.55 to −2.99 | 0.98 to 5.44 | −4.33 to −1.12 | 4.0% | 0.790 (0.406–0.877) |
| | cvi42 4D vs. 2D | 99 | 0.37 (0.05 to 0.74) | 3.12 to −2.37 | 0.92 to 4.57 | −3.14 to −0.86 | 3.9% | 0.853 (0.589–0.915) |
| Total volume (mL) | Slicer vs. 2D | 99 | −2.0 (−2.7 to −1.3) | 3.3 to −7.3 | 1.8 to 5.0 | −9.1 to −5.3 | 21.5% | 0.975 (0.930–0.985) |
| | cvi42 4D vs. 2D | 99 | −1.8 (−2.5 to −1.1) | 3.1 to −6.6 | 1.7 to 4.7 | −8.2 to −4.9 | 20.5% | 0.980 (0.936–0.989) |
| Regurgitation fraction (%) | Slicer vs. 2D | 95 | 0.9 (−0.6 to 2.5) | 17.1 to −15.3 | 11.3 to 22.3 | −18.4 to −10.9 | 3.7† | 0.784 (0.443–0.865) |
| | cvi42 4D vs. 2D | 95 | 0.2 (−1.6 to 2.5) | 15.8 to −15.4 | 10.0 to 21.4 | −18.8 to −11.2 | 3.6† | 0.789 (0.448–0.869) |
| Effective volume per minute (L/min) | Slicer vs. 2D | 100 | −0.20 (−0.28 to −0.12) | 0.33 to −0.72 | 0.19 to 0.45 | −0.87 to −0.56 | 23.2% | 0.956 (0.905–0.972) |
| | cvi42 4D vs. 2D | 100 | −0.18 (−0.26 to −0.11) | 0.28 to −0.65 | 0.15 to 0.40 | −0.76 to −0.52 | 22.4% | 0.966 (0.921–0.980) |
| Peak velocity (cm/s) | Slicer vs. 2D | 100 | −8.1 (−14.5 to −2.0) | 43.0 to −59.3 | 25.6 to 58.2 | −80.8 to −35.1 | 18.8% | 0.777 (0.687–0.894) |
| | cvi42 4D vs. 2D | 100 | 1.7 (−6.4 to 11.3) | 54.4 to −51.0 | 34.6 to 77.7 | −66.7 to −33.2 | 22.9% | 0.809 (0.658–0.903) |
| Heart rate (bpm) | cvi42 4D vs. 2D | 100 | 4.1 (−2.9 to 13.3) | 33.5 to −25.3 | 7.3 to 54.2 | −33.0 to −10.0 | 7.9% | 0.871 (0.715–0.974) |

Bias is the mean difference (4D method minus 2D phase contrast). n is lower for backward and total volume (one ROI without a 2D backward volume) and for regurgitation fraction (five ROIs without a 2D value). *Limits of agreement adjusted for multiple ROIs per patient [42]. MAPE was calculated relative to the 2D value; for backward and total volume, relative to the 2D forward volume. †For regurgitation fraction, the mean absolute difference in percentage points is shown instead of MAPE, because regurgitation fraction can be zero. Effective volume per minute from 4D flow was compared with 2D cardiac output (net volume × heart rate). Heart rate was compared for cvi42 4D flow only. Confidence intervals were estimated by patient-level cluster bootstrap (20,000 resamples). 2D = two-dimensional; 4D = four-dimensional; bpm = beats per minute; CI = confidence interval; ICC = intraclass correlation coefficient; LoA = limits of agreement; MAPE = mean absolute percentage error; ROI = region of interest.

## Appendix A. Native high-dimensional image and transform support in the Insight Toolkit

This appendix is the technical description of the five-dimensional (5D) and higher-dimensional storage capability summarized in the Methods, and the reference for that contribution alongside the source code in ITK and 3D Slicer. Axes are denoted x, y, z (spatial), c (color or channel component), v (vector component), and t (time).

### A.1 Rationale

Time-resolved, multi-component imaging data, such as 4D flow velocity fields (a three-component velocity vector at each voxel at each cardiac phase) and deformable transform sequences (a displacement field at each time point), are 5D: three spatial axes, a component axis, and time. Scanners deliver these data in the Digital Imaging and Communications in Medicine (DICOM) format, which carries rich structured metadata, but each vendor implements it differently, and reading it requires complex parsing and image reconstruction. Research workflows therefore usually import from DICOM once and continue in a simpler container: variants of the Hierarchical Data Format (HDF5), the Neuroimaging Informatics Technology Initiative (NIfTI) format, or the Nearly Raw Raster Data (NRRD) format. Before this work, the NRRD reader and writer of ITK supported only one non-spatial axis, so a 4D flow study or a transform sequence had to be split across many files or handled by application-specific code.

We implemented reading and writing of images, image sequences, vector fields, and transform sequences of five or more dimensions in the NRRD format [33] within ITK [17], the toolkit on which 3D Slicer and many other medical image computing applications are built. NRRD was chosen because it is simple and efficient, its text header has standard fields that state the role of each axis, and it accepts arbitrary custom fields for details the standard does not cover. The feature was implemented in ITK rather than in 3D Slicer, so any ITK-based software, including its Python package, reads and writes these files without custom development; other software can write valid 5D files with modest effort and read specific variants with moderate effort.

### A.2 Supported data types

The following can each be stored in a single NRRD file and loaded for visualization and analysis. Types marked (new) were not previously supported in a single file.

#### 3D

- Volumetric image (xyz).
- 2D image time sequence (xyt).
- 2D color image (cxy).

**4D**

- Color volumetric image (cxyz): cryosection data such as the Visible Human, or multichannel and multimodal images such as B-mode with color Doppler.
- Displacement field (vxyz): a warping transform.
- Flow field (vxyz): one frame of 4D flow magnetic resonance imaging (MRI).
- Sequence of volumetric images (xyzt): echocardiography, computed tomography (CT), MRI, or other dynamic imaging (new).
- Sequence of 2D color images (cxyt): color video (new).

**5D**

- Sequence of color volumetric images (cxyzt) (new).
- Sequence of displacement fields (vxyzt): the result of deformable sequence registration (new).
- Sequence of flow fields (vxyzt): time-resolved 4D flow MRI (new).

Four uses motivated the work. Deformable sequence registration results could previously be saved by the SequenceRegistration extension only as an archive of separate displacement-field files, which was large and gave no random access to a single time point; they are now stored by default as one 5D file, uncompressed for fastest access or compressed for smallest size. 4D flow MRI, previously a set of 4D volumes in an archive, is stored as one 5D vector-field file. Color video and image sequences are stored in one file without the resolution and dimensionality limits of video containers. Multimodality sequences that share a grid and field of view, such as fused B-mode and color Doppler 3D+t echocardiography, are stored together in one file.

### A.3 Reading: mapping NRRD axes to ITK image types

A NRRD file may contain domain axes (space, time) and range axes (list, RGB color, covariant vector, and others) in any order. ITK represents an image either as Image, whose pixel type is fixed at compile time (scalar, RGB, covariant vector, and so on), or as VectorImage, whose number of components is set at run time. To read a NRRD file, domain axes become ITK image axes and range axes become either the pixel component or additional image dimensions; only one range axis can serve as the pixel component, so the axes may need to be reordered. Previously a fixed rule applied: the first non-spatial axis always became the pixel component, so a scalar volume sequence was loaded as a multi-component 3D image and application code had to handle both Image and VectorImage results and sometimes undo the reordering. An AxesReorder property was added to the NrrdImageIO class with three strategies.

- *UseAnyRangeAxisAsPixel* (default, the previous behavior): the first non-spatial axis becomes the pixel component. A non-list range axis (RGB, covariant vector) is read as the pixel of an Image; when only a list axis is present it becomes the pixel of a VectorImage.
- *UseNonListRangeAxisAsPixel*: a list axis (for example time) always becomes an additional image dimension, so that scalar and vector volume sequences load in the same way and application code no longer branches on whether an Image or a VectorImage was produced.
- *UseScalarPixel*: the data are always read into a scalar Image, and every range axis becomes an additional image dimension.

For a file with axes of kind list, domain, domain, domain (in any order), UseAnyRangeAxisAsPixel yields VectorImage<double, 3> over the three domain axes; UseNonListRangeAxisAsPixel and UseScalarPixel yield Image<double, 4> with axes domain, domain, domain, list. For a file with axes list, domain, domain, domain, RGB-color, UseAnyRangeAxisAsPixel and UseNonListRangeAxisAsPixel yield Image<RGBPixel<double>, 4> with axes domain, domain, domain, list, and UseScalarPixel yields Image<double, 5> with the color axis appended as a fifth dimension.

### A.4 Writing higher-dimensional files

The NRRD writer previously handled one component axis and treated every remaining axis as spatial. One axis can now be designated as a list by setting the NRRD_kinds metadata entry of that axis to list, which distinguishes a time axis from the spatial axes. The header of a 5D array with a leading vector component, three spatial (domain) axes, and one list axis reads:

```
dimension: 5
space: left-posterior-superior
sizes: 3 320 320 64 25
space directions: none (-1,0,0) (0,-1,0) (0,0,1) none
kinds: vector domain domain domain list
encoding: gzip
space origin: (0,0,0)
```

*Listing A.1. Key fields of the NRRD header of a 5D vector-field sequence: a three-component vector axis, three spatial axes of 320 × 320 × 64 voxels, and 25 time points.*

Data may be written uncompressed (raw) for fastest access or compressed (gzip) for smallest size. By convention, sequence files use the composite extension .seq.nrrd; the reader does not require it, but it lets users and applications distinguish a sequence from a single 3D volume.

### A.5 Axis units and per-axis metadata

The NRRD input and output now read and write the per-axis units field. Per-axis metadata (kind, label, and units) is preserved for every axis rather than for spatial axes alone, and the pixel component's own metadata is stored under a pixel key instead of an axis index (NRRD_kinds[pixel], NRRD_labels[pixel], NRRD_units[pixel]), so that component-level information survives a read-and-write cycle. The time axis carries its sample times through two custom fields, axis N index type (numeric) and axis N index values (the list of times, with the unit given in the units field, for example seconds), where N is the zero-based index of the list axis; this is how a sequence records non-uniform frame times such as cardiac trigger times.

### A.6 Transforms and displacement fields

With standard NRRD fields alone, a displacement field cannot be distinguished from any other vector field. A file intended as a displacement field carries the custom field intent_code set to 1006, the value the NIfTI specification assigns to displacement fields, so that applications reading displacement fields from either NRRD or NIfTI recognize them by the same code. A vector field that is not a transform, such as a velocity field saved as a vector image, does not carry it.

### A.7 Integration in 3D Slicer

3D Slicer [15] processes images through pipelines of the Visualization Toolkit (VTK), so two wrapper classes, vtkITKImageSequenceReader and vtkITKImageSequenceWriter, expose the ITK reader and writer to VTK. They are used by the volume-sequence and transform-sequence storage nodes (vtkMRMLVolumeSequenceStorageNode and vtkMRMLTransformSequenceStorageNode), so that any supported data type of any dimension is loaded and saved through the standard interface:

```
node = slicer.util.loadNodeFromFile('path/to/file.seq.nrrd')
slicer.util.saveNode(node, 'path/to/file.seq.nrrd')
```

*Listing A.2. Loading and saving any supported data type of any dimension in 3D Slicer.*

A file saved from 3D Slicer records the class of the node it came from in the custom field DataNodeClassName (for example vtkMRMLScalarVolumeNode), so that it is restored as the same kind of node. The 4D flow modules described in this paper hold the velocity field as a sequence of grid transforms and save it as one 5D file through the transform-sequence storage node.

### A.8 Availability and usage

The functionality is available in ITK 6 (including the ITK Python package) and in 3D Slicer 5.10 and later. In ITK Python, a 5D field is read by selecting the axis-reordering strategy, instantiating the image type from the file's component count and dimension, and reading with the standard reader; it is written with the standard writer:

```
import itk
# read: pick the axis-reordering strategy, build the image type from the
file
nrrd_io = itk.NrrdImageIO.New()
nrrd_io.SetAxesReorderToUseNonListRangeAxisAsPixel()
nrrd_io.SetFileName(filename)
nrrd_io.ReadImageInformation()
PixelType = itk.Vector[itk.F, nrrd_io.GetNumberOfComponents()]
ImageType = itk.Image[PixelType, nrrd_io.GetNumberOfDimensions()]
reader = itk.ImageFileReader[ImageType].New()
reader.SetFileName(filename)
reader.SetImageIO(nrrd_io)
reader.Update()
image = reader.GetOutput()
# write with the standard writer
writer = itk.ImageFileWriter[ImageType].New()
writer.SetFileName(output_filename)
writer.SetImageIO(itk.NrrdImageIO.New())
writer.SetInput(image)
writer.Update()
```

*Listing A.3. Reading and writing a 5D vector field in ITK Python.*

Files that follow these conventions can also be written by any NRRD library without ITK. With the pynrrd package, a 4D volume sequence is written by giving the header the axis kinds, the per-axis labels and units, the space directions with a nan row for the time axis, and the two index fields that carry the frame times:

```
import numpy as np
import nrrd
img = np.zeros([201, 201, 201, 2])  # x, y, z, t
header = {
    'space': 'left-posterior-superior',
    'kinds': ['domain', 'domain', 'domain', 'list'],
    'labels': ['', '', '', 'time'],
    'units': ['', '', '', 's'],
    'space origin': [-137.161, -36.806, -309.719],
    'space directions': [[1.953125, 0, 0], [0, 1.953125, 0], [0, 0,
1.953125],
                         [np.nan, np.nan, np.nan]],
    'axis 3 index type': 'numeric',
    'axis 3 index values': '1.2 1.4',
}
nrrd.write('test.seq.nrrd', img, header)
```

*Listing A.4. Writing a volume sequence with two time points (at 1.2 s and 1.4 s) with the pynrrd package.*

The code is illustrative; class and property names follow the ITK 6 application programming interface.

## Appendix B. The slicerheart_4dflow library of preprocessing and quantification methods for four-dimensional (4D) flow magnetic resonance imaging

slicerheart_4dflow is an open-source Python library of common tools for 4D flow preprocessing and quantification, called by the Import 4D Flow and Analyze 4D Flow modules of the SlicerHeart extension for 3D Slicer. The same tools can be used on their own, as a Python library independent of 3D Slicer, and are installable from the Python Package Index (pip install slicerheart_4dflow; within 3D Slicer, slicer.util.pip_install). The library requires Python 3.12 or later and depends only on NumPy and SciPy; VTK is optional for the particle tracer and pathline export, scikit-image for wall-surface extraction, and pynrrd for reading files written by 3D Slicer, including the clinical test cases. It is released under the MIT license. The GitHub repository (https://github.com/JolleyLab/slicerheart_4dflow) is the authoritative source for the methods, their options and defaults, their documentation, and their tests, and should be consulted for current details. This appendix summarizes the methods available at the time of writing (commit 30b389c, 2026-10-21), organized by processing step, with the published methods they are based on when specified. They are initial implementations gathered in one framework so that they can be compared, replaced, refined, and validated by the community, and are not yet extensively validated.

All methods operate on velocity fields in the world coordinate frame of the image. Arrays are indexed [k, j, i] as returned by ITK and 3D Slicer, and image geometry is carried as a 4 × 4 voxel-to-world matrix, so the methods apply without modification to anisotropic, oblique, flipped, and sheared voxel grids in any world coordinate system (left-posterior-superior [LPS] or right-anterior-superior [RAS]). Points, plane centers, normals, and pathlines are world coordinates in millimeters. Anti-aliasing is the one exception: it is applied to the velocity components as encoded, because each encoding direction aliases at its own VENC, and the rotation to world components follows.

### B.1 Preprocessing pipeline

A single function, preprocess, applies the steps in a fixed order: background phase correction, velocity anti-aliasing, noise masking, computation of the phase-contrast magnetic resonance angiogram (PC-MRA), rotation of the encoded components to world coordinates, and, optionally, divergence-free filtering within a lumen segmentation. Background correction precedes anti-aliasing so that unwrapping operates on offset-free phase; the static-tissue fit is therefore made on the aliased field, which static tissue itself does not affect, and the robust fitting methods (B.2) reject aliased candidates. The method for each step is selectable, and the processing time of each step is recorded and reported with the quality report (B.12). By default no divergence-free filter is applied; it runs only when a lumen segmentation is supplied and a method is chosen.

### B.2 Background phase (eddy-current) correction

A low-order polynomial (second order by default, with the order selectable [26]) is fitted to the time-averaged velocity of automatically identified static tissue and subtracted from every cardiac phase [24,25]. Static-tissue candidates are selected in the same way for all fitting methods: the body is separated from air and lung by Otsu's threshold of the time-averaged magnitude [61], which adapts to the air fraction of the field of view (a fixed percentile or value may be given instead, and a field of view without an air class is treated as all body, with a warning); the outer rim of the body is eroded by three voxels to exclude folded-in wrap-around tissue; the quietest 25% of body voxels by temporal velocity standard deviation are retained (a rule also used in commercial software [43]); and the brightest 10% of the PC-MRA within the body and the lumen segmentation dilated by two voxels are removed to exclude vessels. The fitting methods differ in how they reject candidates that are not static tissue, such as slow steady flow and wrap-around:

*ols:* ordinary least squares, the reference fit [24].

*msac:* M-estimate sample consensus, a robust fit in which many minimal-subset fits are scored by a truncated residual cost and the best model is refitted on its inliers, with an inlier threshold of 2.5 times the robust standard deviation of the residuals [62].

*arto:* weighted regularized least squares with automatic rejection of temporally invariant outliers as published [26]: inverse-standard-deviation weights, an l1 penalty on the polynomial coefficients solved by FISTA [63], a first-order fit of the central field of view, and a three-component Gaussian mixture of the weighted residuals with a 3 standard deviation (SD) cut, iterated twice. The published method was validated on 2D phase contrast; its application to volumetric data here is an extension.

*wrls:* this library's adaptation of the Pruitt approach [26], with inverse-variance weights, a ridge penalty, a separate z-score rejection of steady flow, and a two-component mixture with a 2 SD cut, in one pass; retained as an exploratory variant.

*war+msac (default):* MSAC to remove gross outliers, then the wrap-around rejection of wrls on the survivors, then an MSAC refit; chosen as the default because it recovered the imposed offset on the thorax phantom in every scenario tested (B.13).

The fit is made on the time-averaged velocity unless per-frame fitting is requested. The residual static-tissue velocity after correction is reported against the 0.6 cm/s acceptance threshold proposed for regurgitant and shunt quantification [64].

### B.3 Velocity anti-aliasing (phase unwrapping)

Velocities exceeding the VENC (the largest velocity measured without aliasing) wrap around into the encoding range. Spatial unwrapping estimates the Laplacian of the true phase from the wrapped phase [65] and solves the resulting Poisson equation in a single step by discrete cosine transform with Neumann boundaries [66], changing only integer wraps so that measurement noise is preserved; because it acts within a frame, it resolves aliasing that persists throughout the cardiac cycle, as in stenotic jets and surgical baffles. Temporal unwrapping detects jumps larger than VENC between consecutive frames along each voxel's time course [28] and recovers up to three successive wraps in either direction; it assumes that the true velocity changes by less than VENC between consecutive frames, and the wrap-count label map allows voxels where this does not hold to be reviewed and corrected. Both are provided because their relative performance depends on spatial and temporal resolution and noise [27]. Before the spatial solve, the wrapped phase is extended three voxels beyond the noise mask by nearest-neighbor fill so that the step to zero at the wall is not a discontinuity the solve must absorb, which substantially reduces the wraps left by the spatial pass.

*laplacian3d:* Laplacian unwrapping of each frame.

*laplacian4d:* Laplacian unwrapping with time as a fourth dimension, with a Neumann time boundary (default, which does not assume that the frames cover the full cycle) or a periodic one, as used by Loecher et al. for retrospectively gated frames spanning the cardiac cycle [66], so that a systolic peak falling in the first or last frame is unwrapped against its neighbor across the cycle boundary.

*temporal:* jump detection along each voxel's time course.

*laplacian_then_temporal (default):* spatial unwrapping followed by a temporal pass to remove any remaining single-phase jumps.

An alternative Laplacian estimator from wrapped phase differences is also available. The resulting wrap-count image is stored as a label map that can be reviewed and edited with the Segment Editor before the corrected velocities are regenerated.

### B.4 Noise mask and phase-contrast angiogram

Voxels without signal (air, lung) carry random phase and are masked from their low time-averaged magnitude (Otsu's threshold by default; a percentile or fixed value optional) or high temporal velocity standard deviation (above the 97th percentile), with connected components smaller than 50 voxels discarded. The PC-MRA, a velocity-weighted angiogram computed as the root of the time-averaged product of squared magnitude and squared speed, is produced as input for vessel segmentation [67].

### B.5 Divergence-free filtering

Because blood is incompressible, divergence of the measured field within the lumen reflects measurement error. Divergence is computed with a masked finite-difference operator that uses central differences within the lumen and one-sided differences at the wall, exact for affine voxel geometry. Projection methods remove exactly the curl-free component of the noise, which for isotropic noise is one third of its power, and preserve through-plane flow, so they are suitable before quantification; the wavelet method also attenuates the solenoidal component of the noise and is intended for visualization and for gradient-based quantities (B.9 to B.11), not for flow volumes.

*fdm (default when a filter is requested):* finite-difference projection onto divergence-free fields within the lumen, the minimum-norm correction such that the discrete divergence vanishes at every lumen voxel, enforced by conjugate-gradient solution of the normal equations [68,69].

*relaxed:* penalized projection that removes part of the divergence [70]; the default penalty, the square of the smallest voxel spacing, removes about 70% of white-noise divergence, and ten times that value is in effect a hard projection.

*dfw:* divergence-free wavelet denoising in the spirit of Ong et al. [71] and Bostan et al. [72], implemented as a Leray split into divergence-free and curl-free components, an undecimated isotropic wavelet transform of each [73] with normalized convolution so that the lumen wall does not contribute [74], soft thresholding of the divergence-free detail coefficients at 1.5 and the curl-free coefficients at 3.0 times the noise estimate, and recombination. It is not an implementation of Ong's divergence-free wavelet basis, its output is only approximately divergence-free (fdm may be run after it), and fdm is preferred on vessels narrower than about ten voxels, where soft-threshold bias is larger.

*leray:* Fourier-domain Helmholtz-Hodge projection with re-masking, provided as a building block; the periodic Fourier domain does not respect the lumen boundary, so it is not used on its own.

### B.6 Encoding-direction self-check

Vendors order and sign the encoded velocity components differently [13,37], so a permuted or sign-flipped component after import is a silent error. A check developed in this library ranks all 48 signed permutations of the components by the root-mean-square divergence of a low-pass version of the field (time-mean, zero outside the lumen, Gaussian-smoothed, evaluated on the lumen dilated by two voxels): since blood is incompressible, the declared mapping should sit at the minimum, and a wrong mapping turns through-plane flow into a wall-normal layer of apparent sources and sinks. The check reports the ratio of the declared mapping to the minimum (consistent when below 1.5), the discrimination between the minimum and the next distinct mapping, and the best mapping. Open inlet and outlet faces of a segmentation that ends inside the field of view are excluded, and when no distinct mapping is at least 1.5 times the minimum the result is reported as inconclusive rather than as a pass. The check must be run after unwrapping, since wrapped data fail it. A global sign flip ties with the identity and is settled by the flow direction in a vessel with known direction, such as the ascending aorta; a component that is near zero everywhere cannot have its sign tested.

### B.7 Flow quantification and internal consistency

Through-plane flow is computed on user-defined planes or on planes placed automatically at fixed arc-length intervals (10 mm by default, radius 15 mm) perpendicular to a vessel centerline from the Vascular Modeling Toolkit [75]. The plane disc is resampled on a regular grid at half the smallest voxel spacing with trilinear, mask-normalized interpolation of the velocity within the lumen, and a warning is issued when the disc leaves the field of view. Net volume per beat is obtained from the heart rate or R-R interval. Internal consistency is assessed by inflow-outflow balance at user-defined plane sets (for example ascending aorta against arch branches and descending aorta, main pulmonary artery against ascending aorta for Qp:Qs, and caval veins against branch pulmonary arteries in cavopulmonary circulations), reported as percent conservation-of-mass error at each processing stage [3,43]. This is the acceptance test applied to every filter above.

### B.8 Pathlines

Particles emitted from a plane are advected through the time-varying velocity field [76] with a fourth-order Runge-Kutta scheme, trilinear spatial and linear periodic temporal interpolation, and termination where a particle leaves the lumen segmentation. Two tracers are provided: the VTK particle-path filter, the integrator used by ParaView, with the velocity frames linearly interpolated in time by a factor of eight before integration so that the integration step is a fraction of a voxel at peak velocity; and a NumPy reference tracer, with both verified against an analytic orbit. Emission time may be set per seed for staggered release, backward tracing is supported, and output is VTK polydata with per-point time and particle index for animation in 3D Slicer.

### B.9 Derived hemodynamic quantities

From the masked velocity gradient within the lumen (central differences inside, one-sided at the wall) the library computes vorticity and helicity density, with a normalized (relative) helicity option [77,78], kinetic energy density [79], strain rate, and viscous dissipation integrated over the lumen [80], with blood density 1060 kg/m$^3$ and viscosity 3.5 mPa·s as adjustable defaults. Gradients amplify noise: on the aortic CFD field with realistic velocity noise, projection filtering does not reduce the error in these quantities, whereas wavelet denoising reduces it substantially, so they should be computed from dfw-filtered velocity and interpreted with their noise sensitivity in mind.

### B.10 Wall shear stress

Wall points are taken from the marching-cubes surface of the binary lumen segmentation [81], computed with scikit-image when available and otherwise from boundary voxels, and inward normals from the gradient of a smoothed mask, transformed to world coordinates as covectors so that they remain correct on anisotropic, oblique, and sheared grids. Velocity is sampled one, two, and three voxels inward along each normal, the tangential component is fitted with a quadratic through zero at the wall (a linear fit is optional), and wall shear stress is the viscosity times the wall shear rate, with time-averaged wall shear stress and the oscillatory shear index [82] reported [83,84]. On Poiseuille flow the estimate is within 10% at 1 mm voxels on every test geometry. On Womersley flow with a thin boundary layer it underestimates the magnitude at clinical voxel sizes while following the time course; this is the known resolution limit of 4D flow wall shear stress [85] and is recorded by the tests rather than tuned away.

### B.11 Relative pressure

The pressure gradient is computed from the Navier-Stokes equations with temporal, convective, and viscous terms, each switchable [86,87]; the temporal term accepts the frame duration and a non-periodic setting for prospectively triggered data covering part of the R-R interval. Relative pressure is obtained by weighted least-squares integration of the gradient over the lumen, equivalent to the pressure Poisson problem with Neumann boundaries built into the one-sided wall stencils, solved by conjugate gradients [88]; a finite-element formulation has also been described [89]. Uniform weights are the default, because down-weighting the wall layer improves viscous-dominated flow but biases a convection-dominated jet. The gradient and the time course of relative pressure are verified against the Poiseuille and Womersley phantoms. Pressure differences between regions are reported over the cycle [90].

### B.12 Quality report

A per-study report records data quality and the effect of each step, covering items addressed by the consensus statement [3]: VENC adequacy (peak component velocity along each encoding axis relative to VENC, and the aliased fraction from the wrap counts), velocity noise (from the temporal variance of magnitude-selected tissue, rescaled to be unbiased), residual background offset with pass/fail against the 0.6 cm/s threshold [64], lumen divergence before and after processing, noise-mask fraction, the encoding check when a lumen segmentation is given, plane flows and mass-conservation error at each stage [43], and processing time per step. The report is produced as JSON and as a one-page Markdown summary.

### B.13 Validation environment

The library includes a synthetic testing environment for fast, repeatable validation without data download. Phantoms with analytic ground truth include steady Poiseuille flow with exact pressure gradient and wall shear stress; pulsatile flow in a tube with exact velocity, flow, gradient, and wall shear stress per frame [34]; a cosine stenosis with a lubrication-theory velocity field that is exactly divergence-free and flow-conserving, with known throat velocities, centerline pressure, simplified-Bernoulli gradient, and wall shear stress contrast, and a jet exceeding a clinical VENC throughout the cycle [35]; a bifurcation with daughter radii following Murray’s law and a set flow split, returning measurement planes and consistency checks so that mass conservation and Qp:Qs-type ratios have a known answer [36]; and a thorax-like phantom with lungs, a pulsatile artery, a slow steady vein, a second-order background offset, and a phase-encode wrap-around band. A phase-contrast forward model with four-point encoding [37] and complex Gaussian noise produces velocity noise, aliasing, and partial-volume effects (by complex block averaging) from the signal model itself, so that the reference velocity on a downsampled grid is the signal-weighted value the measurement actually sees. An aortic CFD field [38] provides a realistic ground truth and is downloaded on first use from a release of the repository and verified by its hash.

Representative results from this tier: on the thorax phantom the default war+msac background correction recovered an imposed offset field in all scenarios, whereas ordinary least squares erred when wrap-around and steady flow were both present. On the aortic CFD field with added noise, finite-difference projection reduced the velocity error while preserving through-plane flow, and wavelet denoising reduced the velocity and angular errors substantially more, at each of the clinical voxel sizes tested. With a jet exceeding VENC throughout the cycle, Laplacian followed by temporal unwrapping left far fewer residual wrapped voxels than temporal unwrapping alone. Flow is robust to resolution, whereas the oscillatory component of wall shear stress is not recovered above 2 mm on the Womersley phantom and the viscous pressure term degrades at coarser voxels.

Every test runs automatically on each push and pull request through GitHub Actions, on Linux (Python 3.12 to 3.14), Windows, and macOS, and with the VTK version shipped in 3D Slicer. Every recorded metric, 152 at the time of writing, is compared with a committed baseline at a 5% relative tolerance, so that a change in the code that moves a result is detected even when the result stays within its hard limit; processing times are recorded but not compared. Invariance tests confirm that each method is unchanged under world-frame transforms (LPS to RAS, rotation, reflection, translation), velocity and VENC scaling, cardiac-phase rolling, and one-voxel dilation of the segmentation within stated tolerances.

Synthetic data cannot test vendor-specific import and encoding conventions, scanner noise and background-offset characteristics, or agreement with other software, and measured data carry no ground truth. A clinical tier therefore runs on measured studies exported from the Import 4D Flow module as test cases: the magnitude and raw velocity as converted from DICOM; the velocity, wrap counts, PC-MRA, and plane flows produced by SlicerHeart; a lumen segmentation and measurement planes; and, where available, flows reported by other software for the same planes, all in 3D Slicer’s native file formats with a structured acquisition description in place of the DICOM header. The library reads these cases without 3D Slicer or a DICOM reader. For each case the tests check that the files carry no identifiers and describe one consistent pipeline; that the encoding-direction check agrees with the component mapping used at import; that every background-correction method leaves less than 0.6 cm/s in static tissue; that every unwrapping method agrees with the reviewed SlicerHeart wrap counts and gives a physiological peak speed; that VENC adequacy in the quality report agrees with the wraps applied; that plane flow agrees with the SlicerHeart modules and, where a case carries them, with reference measurements from other software; that inflow-outflow balance and flow ratios at the plane sets of each case are recorded and, for planes placed or reviewed by a person, held within tolerance; that the divergence-free filters reduce lumen divergence; that results are unchanged between world frames; that pathlines remain within the lumen; and that derived hemodynamic quantities are in physiological ranges. The tolerances of each check are documented in the repository. Every number is recorded in a clinical regression baseline stored with the data rather than in the repository, one section per case with the hashes of the files it was computed from, so that a re-exported or re-annotated case marks its baseline stale instead of reporting a false regression.

The clinical tier passes on 15 compressed-sensing 4D flow CMR studies acquired at Children's Hospital of Philadelphia on a 1.5 T Siemens scanner. The cases are available in the GitHub testing suite. Support for GE and Philips data is planned as open data becomes available.

### B.14 Assumptions and limitations

These methods are a foundation under active development rather than a finished toolset. They have been tested on synthetic and CFD data with ground truth and on single-vendor clinical data for internal consistency and agreement with the SlicerHeart modules; their agreement with commercial software and their performance across vendors, sequences, and patient populations remain to be established, and community refinement and validation are needed before any of them is relied upon for clinical decisions. Current limitations, each an area of ongoing development, include:

> Scanner-side corrections. Concomitant-gradient (Maxwell) and gradient-nonlinearity corrections are assumed to have been applied by the scanner [91].
>
> Effect on flow measurement. Background correction and anti-aliasing are required for quantification. The projection filters preserve through-plane flow; the wavelet filter alters the flow field and is not intended for flow volumes, and the Fourier projection does not respect the lumen boundary.

Acquisition assumptions. The best unwrapping method depends on resolution and noise, so the default may not suit every acquisition, and temporal unwrapping assumes that the velocity changes by less than VENC between consecutive frames, so its result should be reviewed where systolic acceleration is sharp. Periodic temporal assumptions require frames spanning the full cardiac cycle. The global sign of the velocity components must be confirmed against a vessel with known flow direction.

Resolution limits. Wall shear stress and the viscous pressure term are resolution-limited at clinical voxel sizes (B.10, B.11); the tests record rather than correct this.

Library choices. Several methods and defaults are choices of this library rather than published methods, and methods published for 2D data have not been extensively validated on volumetric 4D flow data beyond the tests described here.

The source code, documentation, and tests are openly available in the repository so that new methods, validation, and refinements can be contributed and evaluated in the same framework.